\documentclass[runningheads]{llncs}
\usepackage{graphicx}

\usepackage{natbib}
\setcitestyle{square,numbers}

\usepackage[toc,page]{appendix}

\usepackage{nameref}

\usepackage{amsmath}
\usepackage{amssymb}
\usepackage[ruled,vlined]{algorithm2e}

\usepackage{enumitem}

\usepackage{xcolor}
\usepackage{color,soul}
\usepackage{pdfpages}

\def\HiLi{\leavevmode\rlap{\hbox to \hsize{\color{yellow!50}\leaders\hrule height .8\baselineskip depth .5ex\hfill}}}
\newcommand{\newpar}[1]{\vspace{-0mm}\paragraph{#1.}}
\definecolor{myyellow}{rgb}{1.0,1.0,0.8}
         
\newcount\Comments  % 0 suppresses notes to selves in text
\definecolor{darkgreen}{rgb}{0,0.6,0}         
         
\newcommand{\kibitz}[2]{\ifnum\Comments=1{\color{#1}{#2}}\fi}
\newcommand{\rmr}[1]{\kibitz{red}{[Reshef says:#1]}}
\newcommand{\rf}[1]{\kibitz{blue}{[Roy says:#1]}}

\begin{document}

\title{Proportional Participatory Budgeting with Projects Interaction}

% \author{Anonymous Submission X}
\author{Roy Fairstein\inst{1} \and
Reshef Meir\inst{2} \and
Kobi Gal\inst{1,3}}
\author{Roy Fairstein\inst{1}\orcidID{0000-0002-2352-5200} \and
Reshef Meir\inst{2}\orcidID{0000-0003-0961-3965} \and
Kobi Gal\inst{1,3}\orcidID{0000-0001-7187-8572}}
%
% \authorrunning{Submission X}
% \institute{}
% \authorrunning{Fairstein et al.}
\institute{Ben-Gurion Univ. of the Negev, Israel \and
Technion, Israel Institute of Technology \and
University of Edinburgh, U.K.}
\maketitle              % typeset the header of the contribution
\begin{abstract}
Participatory budgeting (PB) is a democratic process for allocating funds to projects based on the votes of community members. PB outcomes are commonly evaluated for how they reflect voters' preferences (e.g., social welfare) and the extent to which they are fair (e.g., proportionality).  
Due to practical and computational reasons, voters are usually asked to report their preferences over projects separately, possibly neglecting important dependencies among projects, which causes the outcome to no longer be proportional and achieve lower satisfaction.

This work is the first to suggest a polynomial-time aggregation method capable of guaranteeing proportional outcomes under substitution dependencies. The method is a variant of the Method of Equal Shares, and we further provide another variation that can guarantee a more relaxed notion of proportionality for any type of dependency, and is FPT rather than polynomial. Through simulations, we demonstrate that these aggregation methods achieve, on average, higher social welfare than their counterparts that ignore the dependencies.

\end{abstract}
%
%
%

%%%%%%%%%%%%%%%%%%%%%%%%%%%%%%%%%%%%%%%%%%%%%%%%%%%%%%%%%%%%%%%%%%%%%%%%

\section{Introduction}\label{sec:intro}

Participatory budgeting (PB) is gaining increased attention from both researchers and practitioners and is actively used in cities around the world~\cite{su2017porto, inbook}. 
This process typically includes several steps. In the first step,  citizens suggest and discuss different projects, followed by a stage of shortlisting~\citep{rey2020shortlisting} to get a short list of feasible projects with their cost estimations.
Next,  citizens vote on which of the projects they would like to be funded. Different input formats exist that allow voters to express their preferences~\citep{benade2020preference, fairstein2023participatory} over projects, such as approval voting, knapsack ranking, and additive utility (specified later on). Finally, the votes are aggregated by a mechanism that selects a subset of projects to fund~\citep{talmon2019framework, los2022proportional, peters2021proportional}.

In assessing aggregation methods, various criteria are employed, among which is the notion of proportionality. Proportionality aims to ensure that a group of voters with similar preferences receives a satisfactory level of welfare. One prominent polynomial-time aggregation method, the Method of Equal Shares (ES)~\citep{peters2021proportional}, provides assurance that its outcome will satisfy the proportionality notion of \emph{Extended Justified Representation up to 1 project} (EJR-1), under the assumption that voters' utilities over projects are additive.

In the real world, voters' preferences may exhibit complex relationships between projects. For example,  the utility of one project might depend on whether some other project is funded or not \cite{jain2020participatory, jain2021partition, baumeister2023bounded, goyal2023mechanism}.
Consider, for example, a city with four different suggestions to build a large parking lot in different places, as well as two unrelated projects (say, a playground and a library). The budget is sufficient for only four projects.
There is a severe parking problem, so most citizens will assign higher utility to parking lots than to the other projects (or rank parking lots higher than other projects). 
% However, it is possible that voters might want only 3 of them depending on their location.
As a result, all 4 parking lots are likely to be funded and consume the entire budget, even if one or two parking lots are sufficient to solve most of the parking shortage, and the remaining budget would be put to better use by funding the other projects. This problem occurs since common mechanisms ignore the fact that the four parking lots are \emph{substitutes}.

This example directly shows how ignoring such dependencies may lead to lower welfare.  Furthermore, outcomes that satisfy EJR-1 are no longer guaranteed to exist when considering interactions. We will explain this in detail in Section~\ref{sec:prop}.

\newpar{Our contribution} 

\begin{itemize}
    \item We provide two extensions of ES: \emph{Interaction Equal Shares} (IES) that chooses the next project according to its marginal utility; and \emph{Partition IES} (PIES) that considers subsets of interacting projects at each iteration.

    \item We show that even with few substitute projects, it is no longer guaranteed that an outcome satisfying EJR-1 exists. Therefore, we extend the notion of EJR-1 to consider arbitrary interactions between projects, namely \emph{EJR with Interactions up to 1 project} (EJRI-1).

    \item We prove that IES always returns an EJRI-1 outcome under substitute relations between projects and that PIES holds a more relaxed notion of proportionality for any type of interaction.

    \item We show through experiments that IES achieves higher welfare on average than ES.  \rmr{IES seems consistently higher}\rf{fixed and removed PIES from the statement}
    
\item Our simulations are performed by extending the open source library pabutools~\citep{pabutools} to support aggregation with interaction between projects. This code will be made public for future research.

\end{itemize}

%%%%%%%%%%%%%%%%%%%%%%%%%%%%%%%%%%%%%%%%%%%%%%%%%%%%%%%%%%%%%%%%%%%%%%%%

\subsection{Related Work}

\newpar{Evaluation of aggregation methods}
Past work in PB uses different methods in order to evaluate the voting rules. One method for evaluation that has received a lot of attention is proportionality, see an overview by %having many notions which are surveyed by 
\citet{rey2023computational}.
In our work, we will focus on the notion of Extended Justified Representation (EJR) which will be defined in Section~\ref{sec:prop}.

Another important property of a mechanism is the \emph{social welfare} it provides.  \citet{goel2019knapsack} show that when using the knapsack input format (1/0 additive utilities, i.e., approval), it is possible to achieve an outcome that maximizes social welfare.
While maximizing welfare under combinatorial preferences is generally NP-hard, 
 \citet{jain2020participatory} consider special cases where this is possible. 

 Even putting complexity issues aside, proportional mechanisms do not, in general, maximize welfare~\cite{michorzewski2020price,fairstein2022welfare}. 
%\citet{fairstein2022welfare} studied the tradeoff between welfare, representation and proportionality affects, showing that this requirement can lower the maximal achievable welfare. In addition, \citet{michorzewski2020price} tests the relation between fairness (proportionality) and welfare, 
However, in the settings where projects are \emph{divisible}, stronger positive results can be obtained for projects with unit cost~\citep{brill2023completing}.

\newpar{Projects Interaction in PB}
The literature suggests many different ways to 
represent voter preferences over a discrete set of projects (i.e., input format), such as approval voting \cite{goel2019knapsack, fluschnik2019fair}, knapsack voting \cite{goel2019knapsack, goel2016knapsack}, 
ranking \cite{aziz2020expanding, benade2020preference}, reporting utilities \cite{peters2021proportional} for each of the projects and more. However, all of those methods ignore interactions among projects (such as substitution). 
There are several works which tackle this issue: %~\citep{jain2020participatory, jain2021partition, baumeister2023bounded, goyal2023mechanism}. 
\citet{jain2020participatory, jain2021partition} describe an interaction structure based on a combination of project partitioning and approval voting. This is followed by complexity analysis, for which cases it is possible to find the optimal outcome in terms of welfare. We will adopt this model for our paper.

As we described, there is a variety in the literature that considers the settings where there are project interactions. However, there is a lack of literature that considers proportionality in these settings. The only positive result we are aware of is regarding Fully Justified Representation (FJR). While their paper focuses on additive 
 utilities, \citet{peters2021proportional} mention that the definition of FJR allows for arbitrary interactions and that the Greedy Cohesive Rule (which has exponential runtime)  is guaranteed to find an FJR outcome (in particular, one must exist). 

\citet{goyal2023mechanism} also assumes some partition over the projects with specific types of interactions. For this structure, they suggested a voting rule, which they show is strategyproof under some constraints and finishes with a complexity analysis to find the optimal outcome in terms of welfare under their model. 
In contrast to the previous papers, \citet{durand2024detecting} does not assume any prior knowledge of the relation between projects. Instead, they find and consider project synergies during aggregation, highlighting important properties. However, finding the optimal outcome that considers all synergies is NP-hard, and even approximations limiting synergy size can be impractically slow as size increases.

% In comparison, \citet{baumeister2023bounded} also suggested a model have two main differences, first, they work in multi-winner settings and second, they do not assume a partition over the projects; instead, each voter has their own partitions. They focus on a specific structure of project interaction (which can be considered as a special case of the work of~\citet{jain2020participatory}) and perform an axiomatic analysis of different voting rules for this model. Finally, \citet{durand2024detecting} 
% Considers project synergies during aggregation, highlighting important properties. However, finding the optimal outcome is NP-hard, and even approximations limiting synergy size can be impractically slow as size increases

\newpar{Interactions Outside of PB}
It is worth mentioning that the subject of non-additive or `combinatorial' utilities is discussed much further in the literature than participatory budgeting. \citet{baumeister2023bounded} considers interaction in multi-winner settings (a private case of PB with unit cost for all projects). They let each voter describe their own partitions in some specific structure (which can be considered as a special case of the work of~\citet{jain2020participatory}) and perform an axiomatic analysis of different voting rules for this model. 

Furthermore, while \citet{brill2023proportionality} focuses on multi-issue elections (There are multiple issues and decisions out of several options are required for each one), their approach is similar to ours i.e. we both search for proportionality guarantees that will take interactions into consideration and we both use variations of ES~\citep{peters2021proportional} to achieve such outcomes. The main difference is in the domain, where they need to make a decision for each issue while we need to select a limited number of projects under the budget constraint. This also affects the essence of what it means for issues or projects to depend on each other. Finally, the aggregations in both papers are quite similar, with the main difference being that they consider all sub-outcomes, while we consider only a subset of projects in the same partition, thereby limiting runtime enough to enable actual implementation (and indeed, we implemented and tested it).

 The challenges that we described in the introduction also exist in other areas of economics. For example, in combinatorial auctions under different utility functions~\citet{blumrosen2007combinatorial}, and in fair allocation, where \citet{chaudhury2021fair} study the problem of allocating indivisible goods under subadditive valuations for the items. Decreasing marginal gain is also a common assumption in \emph{continuous} budgeting problems~\cite{fain2016core,wagner2023strategy}.

%%%%%%%%%%%%%%%%%%%%%%%%%%%%%%%%%%%%%%%%%%%%%%%%%%%%%%%%%%%%%%%%%%%%%%%%

\section{Preliminaries}\label{sec:prem}

The model is based on the project interactions described in ~\citep{jain2020participatory}. 
We start by defining the Participatory budgeting  instance as an election tuple $(P,cost,\mathcal{Z},L,V, F)$, where:
\begin{itemize}
    \item $P=\{p_1,\ldots,p_m\}$ is a set of $m$ projects.
    
    \item cost: $P\rightarrow \mathbb{R}_+$ is  specifying the cost of each project $p\in P$. For a subset $T\subseteq P$, we denote $cost(T):=\sum_{p\in T}cost(p)$.
    
    \item $L\in \mathbb{R}_+$ is the total budget. 
    
    \item $\mathcal{Z}=\{Z_1,\ldots,Z_{|\mathcal{Z}|}\}$ is a partition of $P$, where $\cup_{i\in [|\mathcal{Z}|]}Z_i=P$ and $\forall i,j \in [|\mathcal{Z}|]$; $Z_i\cap Z_j=\varnothing$. 

     % \item $V=\{v_1,\ldots,v_n\}$ is a set of n voters, where $v_i$ specifies the set of projects $T\subseteq P$ which are approved by voter $i$.
\end{itemize}
The preferences elicited from voters are composed of two parts:
\begin{itemize}
    \item Each of the $n$ voters $V=\{v_1,\ldots,v_n\}$ submits an approval ballot, where $v_i$ specifies the set of projects $T\subseteq P$ approved by voter $i$.
    
    \item A set of \emph{interaction functions} $F_i=\{f_{iZ_1},\ldots,f_{iZ_{|\mathcal{Z}|}}\}$, where $\forall i\in [n],\forall Z\in \mathcal{Z}; f_{iZ}: \mathbb{N}^+ \rightarrow \mathbb{R}^+$.
\end{itemize}

Given a PB instance, an \emph{aggregation method} will return a feasible outcome, i.e., $W\subseteq P$ with $cost(W)\leq L$.

\newpar{Utilities}
We mark by $f_{iZ}(k)$ the utility that voter $i$ gets from the set  $Z\in \mathcal{Z}$ of interacting projects, when exactly $k$ of which are both approved by the voter and selected. Crucially, the voter does not care which projects in $Z$ are chosen as long as she approves them.

Therefore, the utility that voter $i$ gets from outcome $W$ is: 
% \rmr{we might want to emphasize that in the additive model there is no approval ballot. Here it is used to allow the voter to specify that some items in are not relevant (otherwise all items in $z$ are treated equally).
% Perhaps using the term `Approval' is misleading since it sounds related to approval utilities.}
\begin{equation}
u_i(W):=\sum_{Z\in \mathcal{Z}}f_{iZ}(|Z\cap v_i \cap W|),\label{eq:u}    
\end{equation}

and the \emph{marginal utility} that a voter $i$ gets from adding $p$ to subset of projects $T\subset P$ is denoted by 
\begin{equation}
    u_i(p|T) := u_i(T\cup\{p\}) - u_i(T). \label{eq:marginal_u}
\end{equation}

Throughout the paper, we will consider only interaction functions that satisfy the following:
\begin{enumerate}
    \item $f(0) = 0$. % no utility is received if the voter does not approve any project from a part $z\in Z$ that is funded.
    \item %$\forall k>k', f(k) \geq f(k')$: the interaction function 
    $f$ is non-decreasing.
\end{enumerate}

Where specified, we further assume that all functions $f$ are \emph{concave}, corresponding to projects in the same interaction set $Z$ begin substitutes. 
%In addition to the general model, we will look at a family of \textbf{concave interaction functions} that describe substitution between projects in the same partition,~\footnote{

%We do note that some rules, such as some Thiele methods, can be thought of as maximizing welfare under the assumption that \emph{all} projects are substitutes to some extent.} i.e. $f(k+1) - f(k) \leq f(k) - f(k-1)$.

%Note that the interaction function is defined so that it depends only on how many projects were funded from the same group and is indifferent about which projects are funded.

If we go back to the Example from Section~\ref{sec:intro}, we have a partition where all parking lots are in the same set $Z_1$ and the other projects are each in a separate set. A voter that wants at most 2 parking lots might express his interaction function as $f_{iZ_1}=(1,2,2,2)$, i.e.  $f_{iZ_1}(k)=1$ for $k=1$ and  $f_{iZ_1}(k)=2$ for $k\in [2,4]$.\footnote{A voter might want only some of the parking lots due to their locations; thus, their interactions will be defined only for $k\in[1,3]$ or $k\in[1,2]$.
} \rf{What do think about this?}

Suppose one wants to simplify the voting process. In that case, one can provide default interaction functions, allowing the voters to override them or only submit an approval ballot.%~\footnote{It is possible to further simplify it by using approval voting and induce the interactions from it, for example, as done in Section~\ref{sec:exp}. However, this method requires some strong assumptions on the voter's preferences.} \rmr{I would remove the footnote}
In fact, every Thiele method~\cite{thiele1895om} does exactly that: it maximizes social welfare for some concave function $f$, but assuming a single interaction set $Z=P$.

%%%%%%%%%%%%%%%%%%%%%%%%%%%%%%%%%%%%%%%%%%%%%%%%%%%%%%%%%%%%%%%%%%%%%%%%

\section{Aggregation Algorithms}\label{sec:agg}

Our starting point is the Method of Equal Shares (ES)  introduced by \citet{peters2021proportional}. 
ES is an iterative rule, which starts with ``allocating" each voter an equal share of the budget $\frac{L}{|V|}$,  initializes an empty outcome $W=\varnothing$; then it sequentially adds projects to $W$. At each step, to choose some project $p\in P\setminus W$, each voter needs to pay an amount that is proportional to her utility from the project, but no more than her remaining budget (note that with approval utilities this means only agents that approve the project pay). The total payment should cover the cost of the project. 
 
 Formally, let $b_i(t)$ be the amount of money that voter~$i$ is left with just before iteration~$t$. We say that some project $p\in P$ it qValue is defined by the following:

\begin{equation}\label{eq:afford}
 qValue(p,t):=\min\left\{q \text{ s.t. }\sum_{i\in V}\min\{b_i(t),u_i(p)\cdot q\}\geq cost(p)\right\}   
\end{equation}

Where $u_i(p)$ is the utility of voter $i$ for project $p$. 
If there is a qValue that satisfies Equation~\ref{eq:afford}, we say that project $p$ is q-affordable, and we will denote by $pay_i(p):=min(b_i(t),u_i(p)\cdot q)$, the amount voter $i$ needs to pay for project $p$ if it is funded.

%  We say that some project $p\in P$, is $q$-affordable (we call this value qValue) if $\exists q\in \mathbb{R}_+$ such that 
%  \begin{equation}\label{eq:afford}
%  \sum_{i\in V}min(b_i(t),u_i(p)\cdot q)\geq cost(p)
%  \end{equation}
% \rmr{Can you define it as
% $$qValue(t):=\min\left\{q\geq 0 \text{ s.t. }\sum_{i\in V}\min\{b_i(t),u_i(p)\cdot q\}\geq cost(p)\right\}$$}

% In each iteration, the projects which is affordable with the smallest value of $q$ is chosen. 
If no candidate project is $q$-affordable for any $q$, ES terminates and returns $W$. Otherwise it selects project $p^{(t)}\notin W$ that is $q$-affordable for a minimum $q$, where individual payments are given by $pay_i(p^{(t)})$. We then update the remaining budget to $b_i(t+1):=b_i(t)-pay_i(p^{(t)})$.~\footnote{For the interested reader a pseudo-code for ES can be found in the appendix.}

We emphasize that ES assumes additive utilities and does not consider project interactions (recall the parking lots example from the introduction).

\newpar{Interaction Equal Shares}
We will now extend ES to \emph{Interaction Equal Shares} (IES). During the run of IES, there are no fixed project utilities, but rather \emph{marginal utilities} that are updated with every iteration: in each iteration where the set of projects $B\subseteq P$ was chosen so far, we calculate the qValues (see Eq.~\eqref{eq:afford}) using $u_i(p|B)$ (see Eq.~\ref{eq:marginal_u}) instead of $u_i(p)$. 
Note that both ES and IES run in polynomial time.

The following example shows how each of the mechanisms works on an instance with substitutes. 

% \par{Example}
\begin{example}

% \begin{exmp}
Consider the 
PB scenario  where $V=\{v_1,v_2\}$, $P=\{a,b,b',c,c'\}$. In the following table, the first line describes the interaction functions for each interaction set (both voters have the same functions), followed by the approval sets of each voter. 

\begin{table}[h]
        \centering
        \begin{tabular}{c|c|c|c}
             $\mathcal{Z}$& $Z_1=\{a\}$ & $Z_2=\{b,b'\}$ & $Z_3=\{c,c'\}$\\
             \hline\hline
             $f_Z$ & $(1)$ & $(1, 1.2)$ & $(1, 1.2)$  \\
             \hline
             $v_1 \cap Z$ & $\{a\}$ & $\{b\}$ & $\{c,c'\}$  \\
             \hline
             $v_2 \cap Z$ & $\{a\}$ & $\{b,b'\}$ & $\{c\}$  \\

        \end{tabular}
    \end{table}%\vspace{-4mm}

The budget $L=2$ and  $cost(a)=\frac{11}{10},cost(b)=cost(c)=1, cost(b')=cost(c')=\frac{1}{3}$. In the case that ES is used, it has no knowledge of interactions, so all projects have a utility of 1.

\begin{description}
    \item[ES] Project $a$ is $\frac{11}{20}$-affordable; projects $b$ and $c$ are $\frac{1}{2}$-affordable; projects $b'$ and $c'$ are $\frac{1}{3}$-affordable. ES takes the project with the lowest qValue (using lexicographic tie-breaking), which is $b'$, followed by $c'$,  as values do not change over iterations.  
    
    % Each voter is  left with a budget of $\frac{2}{3}$, and  projects are still $\frac{11}{20}$-affordable for $a$ and $\frac{1}{2}$-affordable for $b$ and $c$. 
    In the next iteration, project $b$ will be chosen, and ES will terminate as there are no q-affordable projects. The final bundle of chosen projects is  $W_{ES}=\{b,b',c'\}$ and the social welfare is $u_{1,Z_2}(1)+u_{1,Z_3}(1)+u_{2,Z_2}(2)=1+1+1.2=3.2$ as voter 2 got two substitute projects. 
    
    \item[IES] IES starts the sane by funding $b'$ and $c'$.
    % This procedure begins the same way as ES as no project was funded yet, and all utilities are 1; therefore, projects $b'$ and $c'$ are funded in the first two iterations. 
    
    Since  project $b$ is substitute for $v_2$ and project $c$ is substitute for $v_1$, their utility changes  to $u_2(b)=u_1(c)=\frac{1}{5}$ in the next iteration, making them $\frac{5}{6}$-affordable. Project $a$ is not substitute to either and stays with the same affordability.
    % Since project $a$ is not affected it stays $\frac{11}{20}$-affordable.

    Therefore, project $a$ with the lowest qValue will be chosen
    % Project $a$ has the lowest qValue. Therefore, it will be chosen
    , and IES will terminate as no item is q-affordable anymore.
    The outcome is $W_{IES}=\{a,b',c'\}$ and the social welfare is $1+1+1+1=4$ (project $a$ provides utility 1 for each voter, project $c'$ provides utility 1 for $v_1$ and project $b'$ provides utility 1 to $v_2$).

\end{description}
\end{example}

As can be seen from the example,  when using  ES, voter 1 gets two substitute projects. In contrast, when using IES, the mechanism will prioritize using voter funds for projects that are not substitutes, even if they are more costly.
A more detailed welfare analysis of the different methods can be found in Section~\ref{sec:analysis}.

%%%%%%%%%%%%%%%%%%%%%%%%%%%%%%%%%%%%%%%%%%%%%%%%%%%%%%%%%%%%%%%%%%%%%%%%

\section{Proportionality}~\label{sec:prop}
% \rmr{I have a hunch but I'm not sure how to formalize it. The ES algorithm is performing greedy optimization of some function (I'm not sure what is the function but it looks like some variant of the welfare constrained by individual preferences and budgets. If we had the optimum of this function, it would (trivially?) hold EJR. In additive setting the greedyness may `cost' us a 1-item approximation, just like greedy knapsack is at most 1 item far from passing the optimal solution. Now in your model I think we get the same property (which is not true for subadditive functions in general!) because within each interaction group the optimal order depends only on costs. If we can formalize this idea we can get a much simpler proof that is also more general.}

In this section, we revisit the definition of EJR-1 from the additive setting~\cite{peters2021proportional}. Then, we show that an outcome that satisfies EJR-1 may not always exist in interaction settings. As a solution, we provide an extension called EJRI-1 suited for such settings. Not surprisingly, ES may not meet EJRI-1 (as it ignores interactions altogether). Finally, we prove that IES satisfies EJRI-1 with substitute interactions and propose a variation of IES that satisfies a more relaxed proportionality notion for all interaction types.

\subsection{Proportionality with Additive Utilities}
Let us recall the proportionality definition for additive utilities~\citep{peters2021proportional} where each voter $i$ gets a utility of $u_i(p)$ for project $p\in P$. For a function $\alpha: P\rightarrow [0,1]$, we mark $\forall T\subseteq P; \alpha(T):=\sum_{p\in T}\alpha(p)$.

\begin{definition}[Cohesive group~\cite{peters2021proportional}]~\label{def:cohesive}
    A group of voters $S$ is $(\alpha,T)$-cohesive, where $\alpha: P\rightarrow [0,1]$ and $T\subseteq P$, if $|S|/n \geq cost(T)/L$ and if $u_i(p)\geq \alpha(p)$ for all $i\in S$ and $p\in T$. 
\end{definition}

\begin{definition}\label{def:ejr}
(Extended Justified Representation \underline{Up To One Project}---EJR\underline{-1}~\cite{peters2021proportional}). A rule $R$ satisfies EJR if for each election instance $E$, each $\alpha: P\rightarrow [0,1]$, $T\subseteq P$, and each $(\alpha, T)$-cohesive group of voters $S$ there exists voter $i\in S$ such that  $u_i(R(E))\geq \alpha(T)$. We say $R$ satisfy EJR-1  if we relax the last condition s.t.  for some $p^* \in T$ it holds $u_i(R(E)\cup\{p^*\})> \alpha(T)$. 
    
\end{definition}

\subsection{Proportionality with Interactions}

We start by demonstrating what happens if one tries to use EJR-1 as is for the settings of interactions.
Consider the simple instance with a single voter that approves two substitute projects $P=Z=\{p_1,p_2\}$ with enough budget to fund both and $f_Z(k)=(1,1)$. 
It is easy to see that the single voter is cohesive for $T=P$ and $\alpha\equiv 1$, so $\alpha(Z)=2$, therefore for $W$ to hold EJR-1 we require that either $u_i(W) \geq \alpha(Z)$ or there is some project $p^*\in Z$ such that $u_i(W\cup\{p^*\}) > \alpha(Z)$. However, even funding both projects (i.e. $W=P$) only provides a total utility of 1---strictly lower than the required utility of $\alpha(Z)=2$. 

This demonstrates an issue where an outcome $W$ can achieve lower utility than the utility that is guaranteed due to it because $\alpha$ is additive while the utility is not. Even if we explicitly disallow a set from blocking itself, there are other instances where no outcome holds EJR-1. For more detailed examples, see the Appendix.
% Appendix~\ref{apx:example}.

% As can be seen in the example, the reason that there might not be an outcome that satisfies EJR-1 is related to Def.~\ref{def:cohesive} where $\alpha$ is only defined for singletons. 
Therefore, we would like to extend the cohesiveness definition such that $\alpha$ will also consider sets of projects, i.e., $\alpha: 2^P \rightarrow R^+$. For some  $p\in P, B\subseteq P$ we will mark the marginal utility according to $\alpha$ of adding $p$ to $B$ by $MU_{\alpha}(p|B):=\alpha(B\cup p) - \alpha(p)$.  In Def.~\ref{def:cohesive_I}, we will use this notion to define EJR with interactions (The difference compared to Def.~\ref{def:cohesive} is highlighted).

\begin{definition}[$(\alpha,T)$-cohesive with interactions] \label{def:cohesive_I}A group $S\subseteq V$ of voters is $(\alpha,T)$-cohesive for a set of projects $T\subseteq P$ and \hl{$\alpha: 2^P \rightarrow R^+$}, if $|S|/n \geq cost(T)/L$ and for any project $p\in T$ and \hl{ a subset of projects $B\subseteq P\setminus\{p\}$ it holds $MU_{\alpha}(p| B) = min_{i\in S}u_i(p| B)$ }. 
\end{definition}

\emph{Extended Justified Representation with interactions up to one project (EJRI-1)} is the same as Def.~\ref{def:ejr}, except we now use Def.~\ref{def:cohesive_I} for cohesive groups.

Under the new definition, we always have at least one outcome that satisfies EJRI-1 (See next section that IES outcomes always satisfy EJRI-1). For example,
going back to the example above, the single voter is still cohesive for the two projects; however, under Def.~\ref{def:cohesive_I},  $\alpha(P)=1 = u_i(p_1)$. Thus, any outcome with at least one project holds EJRI-1.

\subsubsection{Rationale Behind EJRI-1 }

Let's recall a stronger notion of proportionality than EJR that was suggested by \citet{peters2021proportional}, called Fully Justified Representation (FJR). See the Appendix for the formal definition. 

This notion was originally defined for additive utilities but naturally extends to utilities with arbitrary interactions. The main difference compared to EJR is that each $T$-cohesive group should be guaranteed welfare according to the entire set $T$, compared to EJR, which puts a bound per project separately, and the guarantee is built as a sum of those bounds. They show that FJR implies EJR, and propose the Greedy Cohesive Rule (GCR), which satisfies  FJR (regardless of interactions) but runs in exponential time.~\footnote{There are no polynomial-time aggregation methods that are known to guarantee FJR.} 

We note that while FJR naturally extends to interactions, it is no longer true that FJR implies EJR. However, it does imply EJRI.%, and in fact, EJR-1 may not hold for any outcome.

We argue that EJRI-1 is the correct extension for the interaction settings for four reasons (proofs for the top three points can be found in the Appendix):
\begin{itemize}
    \item EJRI $\leftrightarrow$ EJR under additive utilities.
    \item EJR if exists $\rightarrow$ EJRI under substitute interactions
    % \item EJRI coincides with EJR under additive utilities, and EJR, if it exists, implies EJRI under substitute interactions (see proof in the appendix).\rf{I'm not sure this is the right way to say the second part} 
    % A rule \mathcal{R} holds EJRI-1 under additive utilities if and only if it holds EJR-1.

    \item FJR implies EJRI  %(see proof in the appendix)%(see Prop.~\ref{prop:FJR_to_EJR} in the appendix).

    \item It is possible to find an outcome that holds EJRI-1 in polynomial time. In particular,  IES satisfies this (Sec.~\ref{sec:IES_EJR}).
\end{itemize}

% To conclude this section, we note that EJR-1 and EJRI-1 are incomparable. There are instances where the outcome that satisfies EJR-1 does not satisfy EJRI-1 and the other way around. One exception is when we have only substitute interactions, in which case EJR-1 (if it exists) implies EJRI-1. See the appendix for details.
% Prop.~\ref{prop:EJR_to_EJRI} in Appendix~\ref{apx:proofs} and discussion in Appendix~\ref{apx:example} for details.

\subsection{Proportionality of Aggregation with Interactions}\label{sec:IES_EJR}
% \subsection{IES Proportionality}\label{sec:IES_EJR}

In this section, we would like to prove that IES outcomes hold EJRI-1. 
% To do so, we will base our proof on the proof that ES outcomes hold EJR-1 in the additive settings~\citep{peters2021proportional} and show how it be generalized for the interaction settings. 
For this purpose, some of the main components of our proof remain as in the additive case, and others are novel. In the following we will present the proof, followed by explaining which components remain the same. 

% we will present the generalized proof followed by an explanation of the main differences compared to the additive-case proof.

\begin{proposition}
    IES holds EJRI-1 with substitute interactions.
\end{proposition}

\begin{proof}

    Consider some PB instances and a $(\alpha,T)$-cohesive group $S$. 
    % Denote $\alpha := \sum_{p\in T}\alpha(p)$. 
    We must show there is some $i\in S$ with utility at least $\alpha(T)$, or exceed it with one additional project.
    
    The proof starts by defining three separate runs:
    
    \begin{enumerate}[label=(\Alph*)]
        \item IES runs on the original instance with the outcome $W$.
        
        \item IES runs on the original instance, but voters in $S$ have no budget constraints when paying for candidates in $T$.
        
        \item IES runs with only voters $S$ and projects in $T$, unlimited budgets, and the interaction functions are defined such that the following holds:
        $$\forall T'\subseteq T, i\in S; u_i(T') = \alpha(T')$$.

    \end{enumerate}%\vspace{-4mm}

    Note that if at the end of run (B) no voter in $S$ exceeds their initial funds of $\frac{L}{n}$, then the outcome of (B) is the same as (A)
    and all projects in $T$ are selected so we are done. %thus hold the required guaranteed utility.
    
    In the case that some voter $i^*$ pays more than $\frac{L}{n}$ in run B, we consider the iteration in which voter $i^*$ first exceeds her budget, and $p^*$ is selected. 
    Let $W'$ be the set of projects selected before $p^*$ (which are the same in runs A and B).

    \begin{definition}
        (cost-utility function). The function $f_{(B)}(x)$ represents the amount of funds that voter $i^* $ had to spend during the run $(B)$ until receiving a utility of $x$.\footnote{To make the function continuous each two points are connected by a straight line such that the utility of a project is given proportional to how much of its cost has been paid so far.} 
        The function  $f_{(C)}(x)$ is similarly defined for the run (C). 
    \end{definition}

    \begin{proposition}\label{prop:conv}
        $f_{(C)}(x)$ is a convex function.
    \end{proposition}

    \begin{proof}%[Proof of $f_{(C)}(x)$ convexity]
        % \rf{add formal proof for the first sentence}
        As we consider only substitution interaction functions, while the cost of projects is unchanged, it holds that for any two project sets $B,B'\subseteq P$ such that $|B| < |B|'$ it holds:
        $$\forall p\in P\setminus (B\cup B'); \alpha(p|B) = min_i u_i(p|B) \leq min_j u_j(p|B') = \alpha(p|B')$$
        % We note that as we consider only substitution interaction functions, the marginal utility of each project can only decrease between interactions while their cost left unchanged. 
        For this reason, their ratio between cost and utility can only increase. Since IES always takes the project with the lowest ratio and the ratio of the other project can only increase between iterations, the slope of $f_{(C)}(x)$ can only increase.
    \end{proof}

    % This proposition means that the $\frac{\text{cost}}{\text{marginal utility}}$  ratio of projects selected in (C) can only be higher as it progresses.
    For the rest of the proof, we note by $\sigma_{p(B)}$ and $\sigma_{p(C)}$ the slope of project $p$ in run (B) or (C), respectively.

    \begin{figure*}[ht!]
    \begin{center}
    \includegraphics[width=10cm]{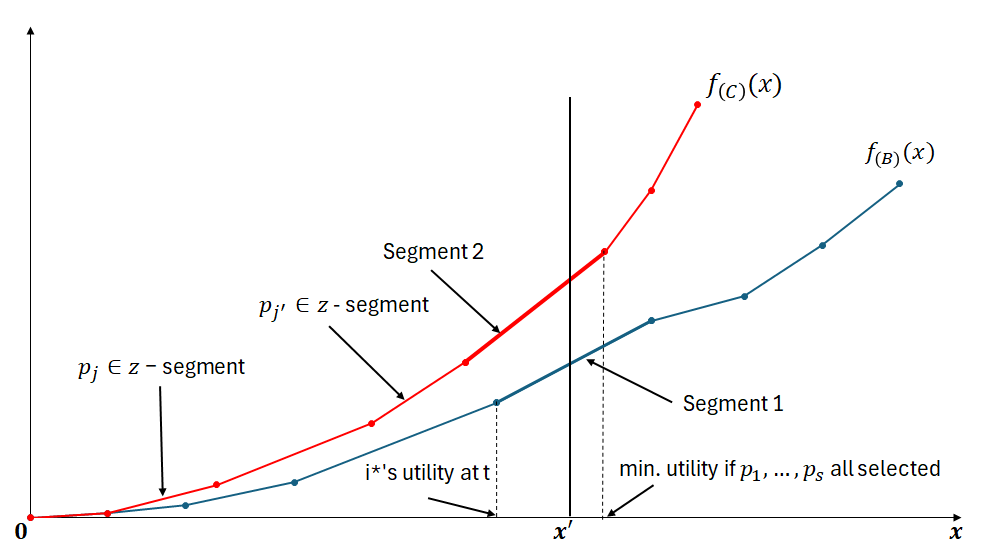}
    \caption{Illustration of the proof of Prop.~\ref{prop:b_lower_c}. X-axis is the utility $u_{i^*}$ received, while the Y-axis is the amount of funds that were used to achieve this utility.    
    Based on \citet[Figure~4]{peters2021proportional}.
    }\label{fig:proof}
    \end{center}
    \end{figure*}

    \begin{proposition}\label{prop:b_lower_c}
        $f_{(B)}(x) \leq f_{(C)}(x)$ for all $x\in[0,\alpha(T)]$.
    \end{proposition}

    \begin{proof}
        Given some point $x'\in[0,\alpha]$ which is not a boundary point for either $f_{(C)}(x)$ or $f_{(B)}(x)$ i.e., where a project is selected (thus having two slopes), we say that $x'$ is on the segment that corresponds to some project $d\in P$ on $f_{(B)}(x)$ (call it segment 1) and on the segment that corresponds to some project $p_s\in T$ on $f_{(C)}(x)$ (call it segment 2). 
        % \rmr{why not segment B and segment C?}\rf{used the same notation is in the original proof. We can change it to B and C}\rmr{I prefer readability to backward-compatibility}\rmr{also, I understand that we have $s=|C_{x'}|$ right? }\rf{yes}
        See Figure~\ref{fig:proof} for an illustration.
        
        Consider the time point $t$ where (B) chooses to add $d$ (before adding it) such that $i^*$ is less than $x'$. We note by $B_{x'}$ the set of projects chosen in (B) until time $t$.
        % utility in (B) is equal to the left $x$-coordinate of the (B) segment, thus less than $x'$. 
        % We note by $B_{x'}$ the set of projects chosen in (B) until getting to utility $x'$, without $d$.
        We define $C_{x'}$ similarly but including $p_s$. Note that by definition $u_{i^*}(B_{x'}) < x' < u_{i^*}(C_{x'})$. 
        
        For a given $Z\in \mathcal{Z}$ denote $z_B:=|B_{x'} \cap Z|$ and $z_C:=|C_{x'} \cap Z|$. We argue that there is some $\hat Z$ s.t. $\hat z_B < \hat z_C$. Let's assume there is no such set, i.e., for all $Z\in \mathcal{Z}$ such that part of it is in $T$, it holds $z_B \geq z_C$. However, as there are no additional projects in $C_{x'}$ it means that $u_i(B_{x'}) \geq u_i(C_{x'})$ in contradiction. 
        
        % and $C_{x'}$ the projects selected by (B) and (C) until getting to utility $x'$ (without $p_s$ and $d$). 
        % \rmr{For a given $z\in Z$ denote $z_B:=|B_{x'} \cap z|$ and $z_C:=|C_{x'} \cap z|$. We argue that there is some $\hat z$ s.t. $\hat z_B\leq \hat z_C$... }\\
        % There is some interaction set $z\cap T \neq \varnothing$ such that $z_B:=|B_{x'} \cap z| \leq z_C:=|C_{x'} \cap z|$.\rmr{so $z_C,z_B$ are numbers while $z$ is a set? confusing}\rf{I tried to define variables to how many projects from z are chosen. Do you have a suggestion for a better notation for this value?}\rmr{My suggestion is only using upper case letters for sets (and only for sets). Use $\mathcal{Z}$ or something for the partition.} If this is not the case, it means that $B_{x'}$ contains more projects from all interaction sets in \rmr{that intersect?} $T$. Therefore, voter $i^*$ should have higher utility in (B) than (C), which is not possible as we look at the same utility point $x'$. 
        % $C_{x'}\subset B_{x'}$, however it is not possible that at point $x'$ (B) have selected all projects as (C) and more as it will have higher utility.
    
        % Therefore, there is some project
        We denote by $p_{j'}\in \hat Z$  the last project in $\hat Z \ \cap\  C_{x'}$ (note that $j'\leq s$).
        % such that $j':= max_k\ k\leq s$.
        As $\hat z_B < \hat z_C$, $\exists p_j\in \hat Z$ with $j\leq j'$ such that $p_j$ was not selected at time $t$ by (B). Therefore, at time $t$ it holds:
        % \vspace{-4mm}
        % At this time, 
        % the utility for $p_j$ in (B) is less than  the utility of $p_{j'}$ when selected in (C) as there is at most the same amount of substitutes from $Z$ at time $t$ in (B). In addition, (C) chooses projects from the same interaction set from cheapest to most expensive (as they share the same utility in each iteration), thus $cost(p_j)\leq cost(p_{j'})$.

        % As a result, the cost-utility ratio of $p_j$ in (B) is at most as the ratio of $p_{j'}$ in (C). Furthermore, from Prop.~\ref{prop:conv}, the ratio of $p_{j'}$ in (C) is at most the ratio of $p_s$ in (C). 
        % Formally, the following holds:

        % \begin{align*}
        %     cost(p_j) \leq cost(p_{j}) % \leq cost(p_s)
        % \end{align*}

        \begin{align*}
            u&_{i^*,(B)}(p_{j}|B_{x'}) = u_{i^*,(B)}(p_{j'}|B_{x'}\cap \hat Z) \\ 
            &=f_{i^*,\hat Z}(\hat z_B+1) - f_{i^*,\hat Z}(\hat z_B)\\
    &\geq f_{i^*,\hat Z}(\hat z_C) - f_{i^*,\hat Z}(\hat z_C - 1) \tag{since $f_{i^*,\hat Z}$ represents substitutes}\\
    &= u_{i^*,(C)}(p_{j}|C_{x'} \cap \hat Z\setminus \{p_j\}) 
        \end{align*}

        From those two inequalities, we get the following:
        
        \begin{align*}
            \sigma_{p_j(B)}& = \frac{cost(p_j)}{u_{i^*,(B)}(p_{j}|B_{x'})} \leq \frac{cost(p_j)}{u_{i^*,(C)}(p_{j}|C_{x'} \cap \hat Z\setminus \{p_j\})}  \\ &=\sigma_{p_j(C)} \leq \sigma_{p_{j'}(C)} \leq \sigma_{p_s(C)} \tag{since $f_C(x)$ is convex}
        \end{align*}
        
        Since (B) always chooses projects with the smallest ratio it must be that $d$ has a ratio at most as this of $p_j$ in (B). Accordingly, this means that $d$ also has a ratio at most as the ratio of $p_s$ in (C).
        Therefore, segment1 has a weakly lower slope than segment2. As it is true for any $x'$ (except boundary points), the proposition holds.

        % Since (B) always chooses a project that is $\rho$-affordable with the smallest $\rho$ it must be that $d$ is $\rho$-affordable with $\rho\leq\rho'$. This means that the slope at (B) segment is at most $\rho$ and therefore at most $\sigma_{p_j}$. On the other hand, the segment on (C) has a slope of $p_s$, which is weakly lower than $p_j$. Therefore, the (B) segment has a weakly lower slope than (C). As it is true for any $x'$ (except boundary points), the proposition holds.
    \end{proof}

    Given the point $\alpha(T)$ and the order the projects are chosen in (C) $\{p_1,\ldots,p_T\}$ it holds:
    $$f_{(C)}(\alpha(T)) = \sum_{i=1}^T \sigma_{p(C)}\cdot MU_\alpha(p_i| \cup_{j=1}^{i-1}p_j) = \sum_{p\in T}\frac{cost(p)}{|S|} = \frac{cost(T)}{|S|} \leq \frac{L}{n}$$
    
    And by Prop.~\ref{prop:b_lower_c} it holds: 
    \begin{equation}\label{eq:b_lim}
        f_{(B)}(\alpha)\leq \frac{L}{n}
    \end{equation}
    
    To conclude, consider the state exactly after (B) adds project $p^*$ where voter $i^*$ hast spent at least $\frac{L}{n}$ at this point. There are two cases:
    
    \begin{enumerate}
        \item $i^*$ spent exactly  $\frac{L}{n}$. In this case $p^* $ is also selected by (A) as (A) and (B) behave the same until this point. According to Eq.~\eqref{eq:b_lim} $u_{i^*}(W_{(B)} \cup {p^*}) \geq \alpha(T)$, thus $u_{i^*}(W)\geq \alpha(T)$.
    
        \item $i^*$ spent strictly more than $\frac{L}{n}$. In this case, by definition of (B), we have $p^*\in T$ such that Eq.~\eqref{eq:b_lim} implies $u_{i^*}(W_{(B)} \cup {p^*}) > \alpha(T)$. As $W_{(B)}\subseteq W$, it implies $u_{i^*}(W\cup {p^*})\geq \alpha(T)$.
    \end{enumerate}

In both cases, $W$ satisfies EJRI-1.

\end{proof}

We highlight two points where our proof differs from the additive-case proof.
% This proof is generalized for the interaction settings in two manners. 
First, in run (C), we define the utilities according to all subsets compared to single projects in the additive case, i.e., $u_i(p)=\alpha(p)$. Second, to prove Prop.~\ref{prop:b_lower_c}, we can no longer use the fact that slopes of non-yet-selected projects in run (B) are bounded by their slope in run (C), which only holds in the additive case where slopes are fixed throughout the run.
% in the additive case, they use the fact that under run (B) at each step, the slope of any not-yet-selected project is at most as the slope of this project when selected in (C). This is only correct in the additive case since the slopes are not affected by the other selected projects until this point.

While we proved that IES holds EJRI-1 for preferences with substitutes, this is not true anymore for the general case,  
which can also be seen in the following example:

\begin{example}\label{exmp:ttif}
    Given PB scenario with 3 voters 
    % $\{v_1,v_2,v_3\}$ 
    and 5 projects 
    % $\{p_1,p_2,p_3,p_4,p_5\}$ 
    where $cost(p_3)=cost(p_4)=cost(p_5)=1, cost(p_1)=cost(p_2)=1.5$, the budget $L=3$ and the voters preference shown in the following table:

    % \vspace{-3mm}
    \begin{table}[h]
        \centering
        \begin{tabular}{c|c|c|c|c}
             & $\{p_1,p_2\}$ & $\{p_3\}$ & $\{p_4\}$ & $\{p_5\}$\\
             \hline
             $f$ & $(0, 10)$ & $(1)$ & $(1)$ & (1)  \\
             \hline
             $v_1$ & $\{p_1, p_2\}$ & $\{p_3\}$ & $\varnothing$  & $\varnothing$ \\
             \hline
             $v_2$ & $\{p_1, p_2\}$ & $\varnothing$ & $\{p_4\}$  & $\varnothing$ \\
             \hline
             $v_3$ & $\{p_1, p_2\}$ & $\varnothing$ & $\varnothing$  & $\{p_5\}$ \\

        \end{tabular}
    \end{table}%\vspace{-3mm}

    Since the utility for the first project from $p_1$ and $p_2$ is zero, neither of those projects will ever be selected by IES. Instead, projects $\{p_3,p_4,p_5\}$ will be selected which
    results in utility 1 per voter.

    Note that all voters are cohesive over the project set $\{p_1,p_2\}$  and should guarantee a utility of 10 to at least one voter. Adding another project will add zero utility, which would not guarantee EJRI-1 either.
\end{example}

To support proportionality for general interactions IES can be modified to consider all subsets of each $Z\in \mathcal{Z}$ at each iteration, we will call this method Partition IES (PIES) and it is defined more formally in the Appendix.%~\ref{apx:pies}. 
This has two drawbacks: one is that runtime is exponential in $|Z^*|:=\max_{Z\in \mathcal{Z}}|Z|$; and the other is that the obtained proportionality notion is further relaxed to EJRI-z instead of EJRI-$1$, i.e. instead of requiring EJRI up to one project $p\in T$, we require EJRI up to one interaction set $Z\in \mathcal{Z};\ Z\cap T \neq \varnothing$ (see the Appendix %~\ref{apx:pies}
for proportionality proof).
While PIES guarantee is weaker, 
% As can be seen, PIES guarantees EJRI-z, which is weaker compared to EJRI-1, promising a cohesive group a lower welfare guarantee. 
 if all interaction sets are relatively small, the guarantees are not much weaker compared to EJRI-1.

Given the results so far, one might ask, what happens if we relax the model such that each voter defines its own partition instead of a single partition $Z$? Regrettably, IES does not hold EJR-1 or EJR-z in this scenario. More details about it can be found in the Appendix/%~\ref{apx:multi_part}.
% \vspace{-1.5mm}

%%%%%%%%%%%%%%%%%%%%%%%%%%%%%%%%%%%%%%%%%%%%%%%%%%%%%%%%%%%%%%%%%%%%%%%%

\section{Welfare Analysis}\label{sec:analysis}

So far we saw under what conditions IES and PIES are proportional, however one of the main purposes of extending the way to express preferences is to allow for higher social welfare. For this reason, in this section, we will focus on analyzing the social welfare that can be achieved by the different methods.
% \vspace{-3mm}

\subsection{Worst-Case Analysis}\label{sec:worst}
% \rmr{I understand that since ES is greedy it fails to optimize welfare, and so do its variants, and due to wrong order or other issues, we have that `by chance' some variants may succeed more than others, with or with out dependencies. 

% I wonder (following the lab discussion) if there is some simple class of instances where ES is (trivially?) optimal for additive instances. Maybe even by severely restricting the costs/preferences/number of participants etc.  If we can show theoretically that on this class IES beats ES on subadditive instances (maybe even IES is optimal) then this would add substantially to our argument.
% }\rf{It is very tricky to show something like this for ES. ES and its variants can be thought of as having two stages, first guaranteeing proportionality and then deciding how to use the rest of the budget (this is usually the part where ES stops but does not exhaust the budget). The second part usually will be the one to control the final welfare.

% In our case, I exhaust the budget using their recommendation: increasing the budget until ES exceeds the budget, and if it does not exceed twice the original budget, we take this outcome and run proportional greedy with the rest of the budget.

% This technique make it difficult to find a class of instances is optimal.
% }

We remind the reader that in the approval settings (without interactions), it was shown that the welfare ratio between ES and the optimal welfare is bounded by the number of voters, and bounds remain similar for all proportional rules~\citep{fairstein2022welfare}. In contrast, this ratio can be much worse when moving to additive utilities. This is because ES can choose the project with the lowest utility while the optimal outcome will take the projects with the highest utility. Since this can happen for all voters between each pair of projects, we get that ES can get a welfare ratio as bad as $\Omega(|P|\cdot|V|\cdot\frac{\text{max utility}}{\text{min utility}})$. \rmr{You mean $\Omega$ not $O$}\rf{fixed}

Since utilities with interactions only generalize additive utilities further, they are susceptible to similar issues. However, getting a worse welfare ratio than the one described  is impossible (i.e., the bound is tight), as every voter can have at most $O(|P|)$ projects, which it gets with the minimum utility instead of the maximal utility. For this reason, the worst-case welfare approximation bounds of ES and IES are equally bad. See the Appendix%~\ref{apx:welfare}
  for details.

While all variations have similar worst-case guarantees, for specific instances, one rule might be better than others. 
Next, we would like to answer whether one of the aggregation methods, ES / IES / PIES, is superior to the other in terms of welfare. To answer this, we start with the following example: \rmr{this is not a substitutes example. Is there a simple example with substitutes?}\rf{Do we want to mention the example from the mail? I'm not sure how much it will add}

 Given partition $Z=\{z_1,z_2\}$ such that each includes half of the projects, the interaction functions for all voters are the same: \begin{equation}
  f_{iz_1}(k)=\sum_{j=1}^k\begin{cases}
    1, & \text{if $j=1$}.\\
    5, & \text{otherwise}.
  \end{cases}\quad ; \quad
  f_{iz_2}(k)=\sum_{j=1}^k 5^{j-1}
\end{equation}

For some small $\epsilon>0$, we define the cost of one project in $z_1$ to be $1-\epsilon$ and the cost for the rest of the projects in $z_1$ to $1 + \epsilon$. All projects in $z_2$ have cost of $1$. The budget is $L=\frac{|P|}{2}$.% for this instance is $L=|z_1|= |z_2|$.

All projects "appear" to ES as with a utility of 1, so it will choose them by their price, i.e., will take the cheap project from $z_1$ and $L - 1$ projects from $z_2$ having the utility of $1 + f_{iz_2}(L - 1)$. In contrast, we have IES, which will also start with the cheap project from $z_1$, but after that, the utility for the rest of the projects in $z_1$ updates to 5, thus IES fund all projects from $z_1$ 
%to use the rest of the budget for only projects from $z_1$, 
with welfare of $f_{iz_1}(L - 1)$. As can be seen, as the number of projects increases, the ratio between ES and IES also increases.

This example shows ES welfare can be significantly higher than IES's, but switching between $z_1$ and $z_2$ costs can reverse this, making ES welfare significantly lower.

% While this example shows that ES welfare can be significantly higher than IES's, we can switch between their outcomes by switching between $z_1$ and $z_2$ costs, thus having ES have significantly lower welfare.

Therefore, we see here that each of ES and IES can be much better or worse than the other for two almost identical instances.

PIES, on the other hand, chooses the optimal outcome for both cases. However, there are scenarios where PIES can have significantly lower welfare than ES and IES. We will see this in the following example:

% \begin{example}
    Given PB instance with three projects, one voter that approves all projects and a budget $L\geq4$. The interaction functions are:
    % \vspace{-1.5mm}
    \begin{table}[!h]
        \centering
        \begin{tabular}{c|c|c}
             & $p_1, p_2$ & $p_3$\\
             \hline
             $f$ & $1, 3$ & $L$ \\

        \end{tabular}
    \end{table}%\vspace{-3mm}

$cost(p_1) = cost(p_2) = 1.1; cost(p_3) = L-1$.

When running PIES, we consider $p_1/p_2$ with 1.1-affordability, $p_3$ with $\frac{L-1}{L}$-affordability and $\{p_1,p_2\}$ with $\frac{11}{15}$-affordability. Since $L\geq 4$, $\{p_1,p_2\}$ will be selected, and PIES will stop. In contrast, ES and IES, considering only one project at a time, will fund $p_3$, followed by one of the other projects, and stop. This means that PIES will stop with a utility of 3,~\footnote{We can also add additional projects that cost $L-2.2$ with utility one so PIES will exhaust its budget} while ES and IES stop with a utility of $L+1$. Therefore, PIES welfare can be significantly lower than the welfare of ES and IES.
% \end{example}

To conclude this section, we saw that in the project interaction settings, neither ES, IES, or PIES dominates the other, and there are instances when each method can be better than the other. 

\subsection{Average Case Analysis}\label{sec:exp}

In the previous section, we examined worst-case scenarios for the aggregation methods, which involve rare extremes. Here, we aim to understand their average behavior across various instances.

% In the previous section, we examined the worst-case scenarios for the aggregation methods. However, those scenarios usually involve extreme cases that are unlikely to occur. Therefore, in this section, we would like to understand how they behave on average in various instances.

The challenge is that, to our knowledge, no instances exist where people express interaction preferences during elicitation. To address this, we will use real-life PB instances from \url{Pabulib.org}~\citep{stolicki2020pabulib}  with approval votes and generate interaction preferences. Additionally, we will create synthetic instances with various possible scenarios and preferences for more comprehensive results.

% The challenge here is that, as far as we know, there are no existing instances where people could express interaction preferences in the elicitation. For this purpose, we will take real-life PB instances from \url{Pabulib.org}~\citep{stolicki2020pabulib} with approval votes and create from them interaction preferences. In addition, to get more complete results, we will create synthetic instances with various possible instances and preferences. 

\subsubsection{Experiment Description}

\newpar{Pabulib Data}
To create the interaction preferences from approval votes, we will use a somewhat similar method as \citet{durand2024detecting}. In their work, they suggest finding synergies between projects by comparing every possible set of projects. However, this is not feasible. Instead, to find the synergy between projects, we will compare the similarity between their approval vectors. The assumption used for this is that projects with very dissimilar votes (having different voters approving them) can be considered substitutes. In contrast, similar projects (voters always choose both or none of them) can be considered complementary.

In practice, projects will rarely be exactly the same or exactly the opposite. Therefore, we will use the distribution of project similarities and choose a percentile that will help create the groups. For example, we can look at projects whose similarity is in the top highest 5\% and top lowest 5\%. Then, we will cluster the projects into groups such that a group is considered complementary if all pairs of projects have similarity in the top 5\%.
% any pair of projects in it have similarity in the top 5\%. 
Similarly, for the substitute groups. By controlling this value, we can analyze how the methods behave when allowing different amounts of interactions, i.e., how many groups there are and how big they can be.
% we enable more projects to interact with each other.

We defined for each group an interaction function as follows:
\begin{itemize}
    \item If the projects are complementary, the marginal utility increases linearly, formally: $f(i) = \sum_{j=1}^i j$.

    \item If the projects are substitutes, we define the utility such that the voter prefers one non-substitute project over the entire set of substitute projects, i.e., the utility for the first project is 1, and any additional project gives the utility of $1/|P|$.
\end{itemize}

\rmr{So did we drop completely the idea of trying to cluster based on description? Also, say not just where is it different from \cite{durand2024detecting} but also where it is conceptually similar.}\rf{Yes, we said we'll leave it for future research.}\rf{Doesn't the "Pabulib Data" paragraph does that? we say there we follow the same idea of synergies but calculate them differently using similarity.}
\newpar{Synthetic Data}
To have more complete results, we would like to experiment with synthetic data in addition. 
% The results so far give us insights into how the different methods behave when controlling the amount of interactions we allow to be between projects. We also did experiments on synthetic data to have a broader image and understand the effect of other factors.
In this experiment, we created instances that include 50 projects with their cost sampled from Normal distribution $N(300,30)$ and split randomly into partitions, each of size $[1,10]$. Next, 100 voters approve randomly 3 partitions, each having interactions as described above (either increasing linearly or decreasing to $1/|P|$). We repeat this 1000 times, and the average results are presented.

The main focus of this experiment is the budget. We want to show how the different methods behave when having different budgets. We expect that when having a small budget, the interaction effect will be small as not many projects can be funded. However, as we increase the budget, it requires to "understand" which interaction set possesses the highest potential, hereby requiring IES to do some "exploration", which is possible as more projects can be funded. An example of such a scenario was presented in Section~\ref{sec:worst}.

\newpar{Experiment Evaluation}
To test our methods, we compared the average welfare the voters receive for four aggregation rules: ES (which does not take interactions into account), IES, PIES, and proportional greedy (greedy by $utility/cost$). The reason for these rules is to see if taking interactions into account can help to achieve better welfare and having proportional greedy (PG) as a baseline, which is known to achieve good welfare.\footnote{While PG is known for achieving high welfare, \citet{fairstein2022welfare} shows that even in the additive scenario, maximizing welfare can come at the cost of proportionality.}

To perform the experiments Pabutools~\citep{pabutools} library was used. As the library currently supports additive utilities but has no interactions between projects, we extended it to support these settings.

\subsubsection{Experiment Results}

\begin{figure}[ht!]
\begin{center}
\includegraphics[width=8.5cm]{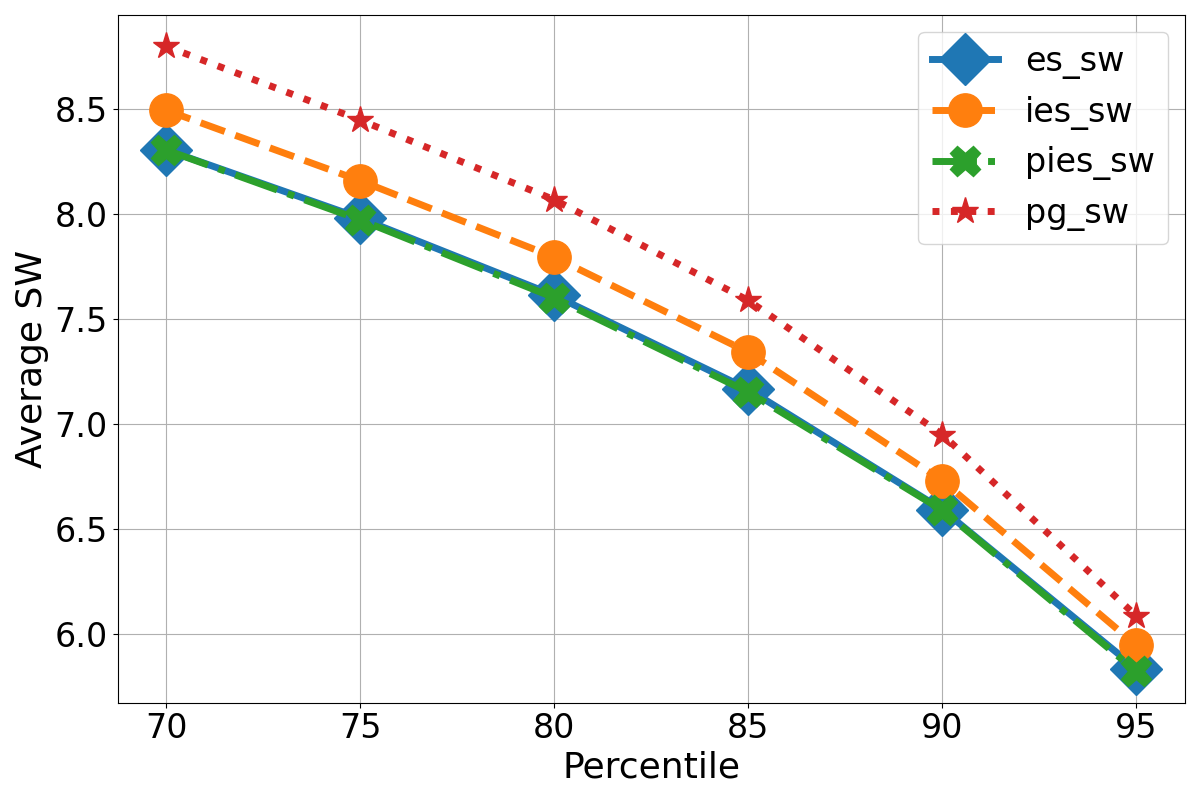}
\includegraphics[width=8.5cm]{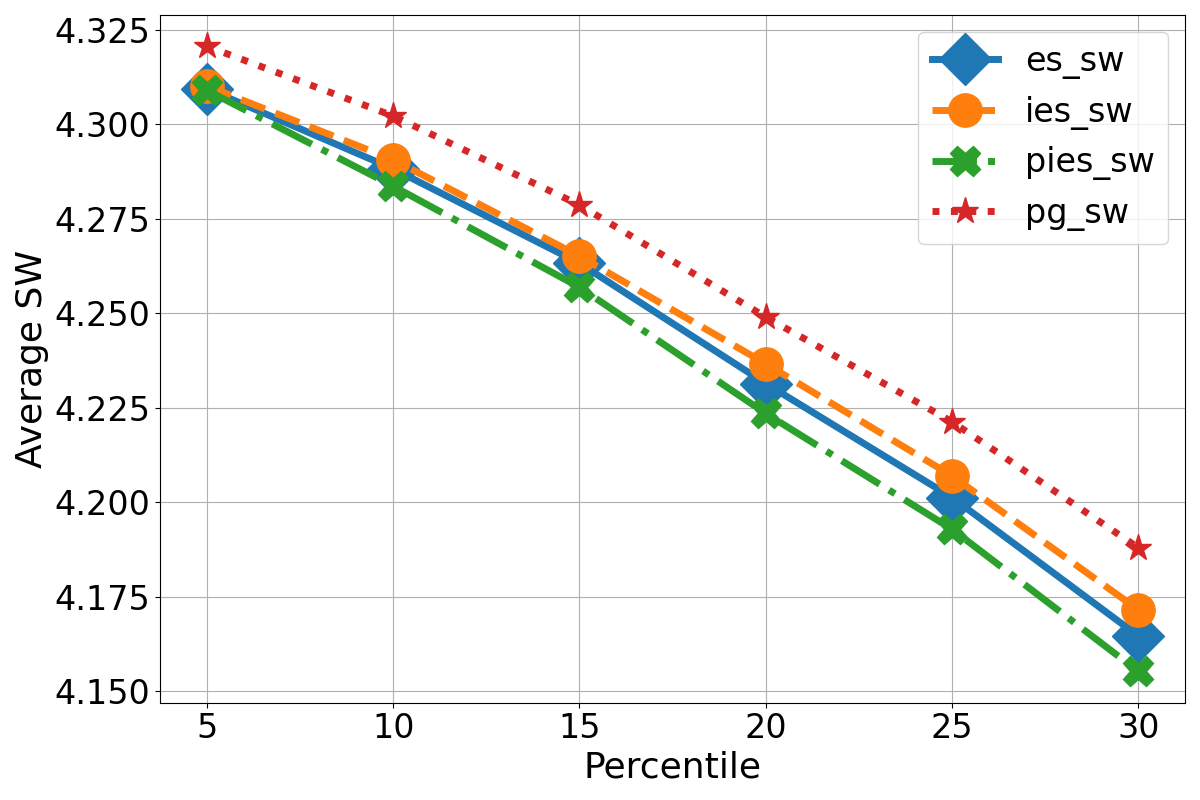}
\caption{Average voter welfare over pabulib data with different percentile for complementary interactions (top) and substitute interactions (bottom).
}\label{fig:avg_sw}
\end{center}
\end{figure}

\begin{figure}[ht!]
\begin{center}
\includegraphics[width=8.5cm]{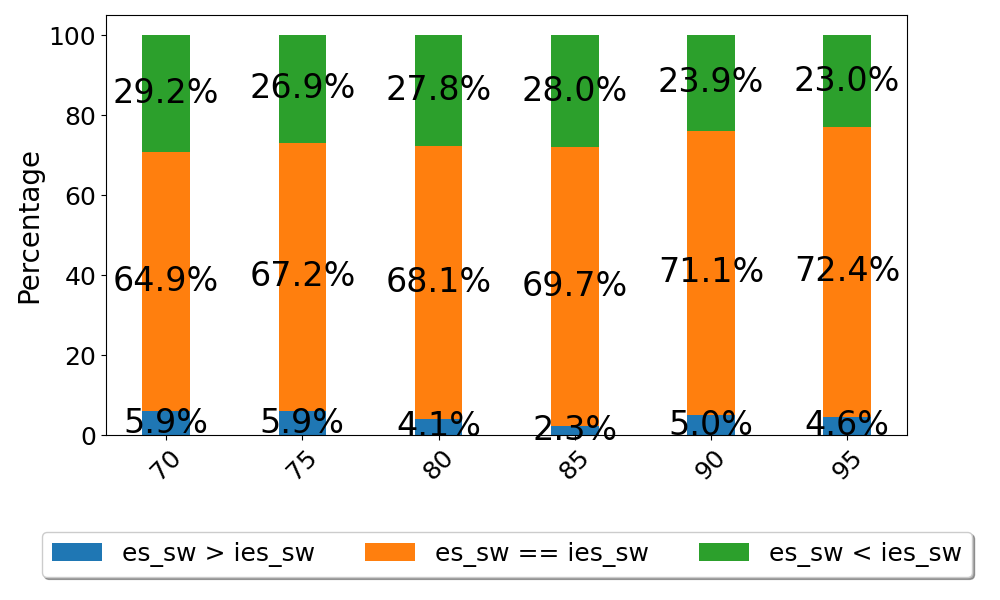}
\includegraphics[width=8.5cm]{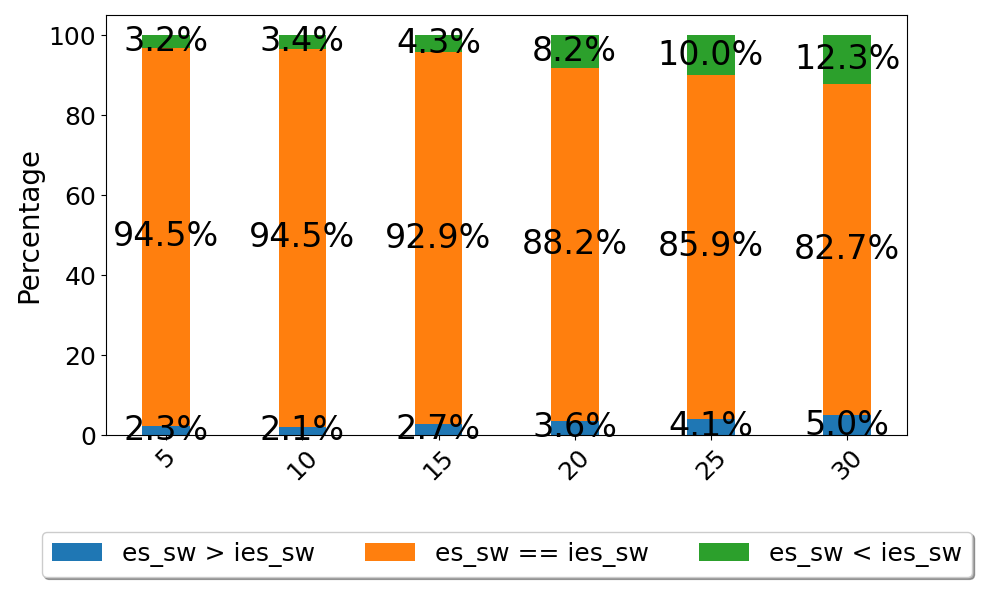}
\caption{Percentage of pabulib instances where IES achieve higher welfare compared to ES. Using percentile 70 for complementary interactions (top) and percentile 30 for substitute interactions (bottom). 
%i.e. extreme complementary, linear and exponential comp .
}\label{fig:percent}
\end{center}
% \vspace{-5mm}
\end{figure}

\newpar{Pabulib Results}
Figure~\ref{fig:avg_sw} presents the main results, showing the average social welfare over all instances when using different percentile values to create the interactions (there are more complementary projects as the percentile decreases and the other way around for substitutes). In both scenarios, we see that, as expected, PG succeeds in achieving the highest welfare; however, it does not give any guarantee about proportionality. Next, we see that by using IES, we succeed in achieving higher welfare compared to ES on average. Furthermore, as we allow more interactions, the gap between their welfare is getting bigger. While the results here are for scenarios where we have substitute or complementary projects, we see similar behavior when having both. %More details about results when having both types of interactions can be found in Appendix~\ref{apx:sim}

% We see that, on average, IES succeed in achieving higher welfare compared to ES when having interactions. 
To complete the picture, Figure~\ref{fig:percent} shows the percentage of instances where IES or ES achieves higher welfare than the other. We can see in the figure that, as shown in Section~\ref{sec:analysis}, neither method dominates the other.
% neither method succeeds in always having at least as high welfare as the other. 
As one could expect, when having a small number of interactions, both methods result in the same welfare in most instances as the interactions have a very small effect (We note that this effect is smaller for complementary projects, as even a few projects can significantly impact outcome welfare). 

Starting with the results for complementary interactions, we see that the percentage of instances in which ES achieves higher welfare compared to IES stays almost the same with any number of interactions, while the percentage of instances in which IES achieves higher welfare keeps increasing as we allow more interactions. In comparison, for substitute interactions, we see that the percentage increases both for ES and IES; however, it increases at a higher rate for IES than ES (IES increases from 3.9\% to 14.1\% while ES increases from 2.1\% to 4.6\%).

\begin{figure}[ht!]
% \vspace{-2mm}
\begin{center}
\includegraphics[width=8.5cm]{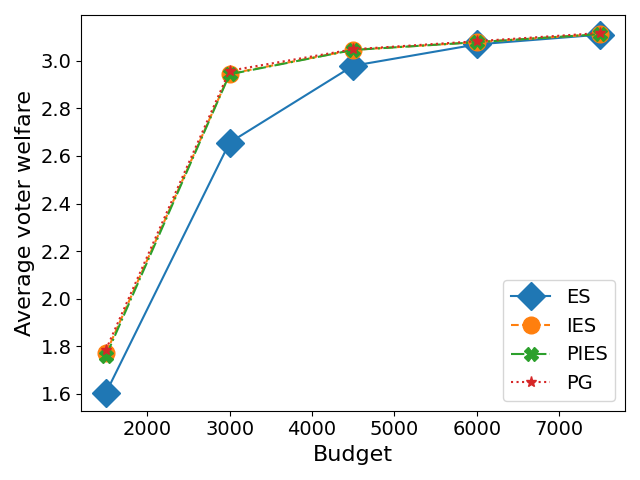}
\includegraphics[width=8.5cm]{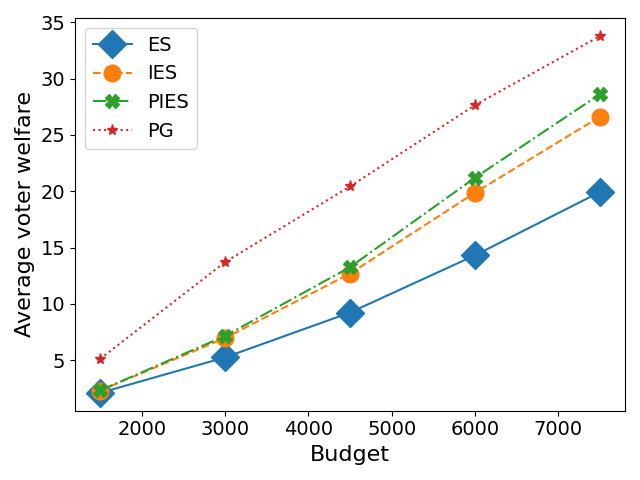}
\caption{Average voter welfare over 1000 simulations, using only substitute interactions (top) or complementary interactions (bottom).
}\label{fig:exp}
\end{center}
\end{figure}

\newpar{Simulation Results}
Figure~\ref{fig:exp} shows that with a limited budget and few fundable projects, the welfare difference between ES and IES is minimal due to limited interaction effects on utilities. As the budget grows, the welfare gap widens, reflecting increased interaction influence on project selection, but this gap narrows as nearly all projects become fundable.

% Observing Figure~\ref{fig:exp}, we note that with a limited budget and few projects fundable, the difference in welfare between ES and IES is minimal, as interactions exert minimal influence on utilities. However, as the budget increases, the welfare gap widens, indicating a greater impact of interactions on project selection. Nonetheless, this gap is expected to narrow as nearly all projects become fundable.

In the substitute scenario, IES achieves welfare levels nearly equivalent to PG, while in the complementary scenario, it outperforms ES but falls short of PG. Despite PG and IES prioritizing project selection based on the cost-to-utility ratio, IES focuses on ensuring fairness over maximizing welfare. 
% Notably, in this scenario, PIES achieves higher welfare by considering sets of projects in each iteration.

While IES looks promising, it is more complex with PIES. Looking at both Figure~\ref{fig:avg_sw} and Figure~\ref{fig:exp} we see that PIES welfare depends on the scenario. While being able to select multiple projects simultaneously is its advantage, it can also be its disadvantage as the mistakes it makes are bigger. For this reason, we see in four different scenarios that  PIES is better than IES, equal to it, equal to ES, and worse than ES. Thus, PIES results with a much higher variance in the welfare it can get.

Finally, we ran additional simulations for different interaction types. The results, included in the Appendix, showed the same overall trends as in this section.

% Finally, we performed more simulations for more types of interactions. The results, which are included in the Appendix, showed the same general trends as those in this section.

%%%%%%%%%%%%%%%%%%%%%%%%%%%%%%%%%%%%%%%%%%%%%%%%%%%%%%%%%%%%%%%%%%%%%%%%

\section{Conclusion And Future Work}

In this paper, we considered proportionality in participatory budgeting in the setting where there is interaction between projects.  We defined a variation of the  EJR/EJR-1 proportionality axiom that considers interactions and suggested two variations of the celebrated ES mechanism that are proportional under project interactions.  The first (IES) runs in polynomial time and guarantees proportionality under substitutes. The second (PIES) is approximately-proportional under arbitrary interactions, but its runtime may exponentially depend on the size of the interaction sets.
% , and approximation guarantees similarly deteriorate. 

We extended pabutools~\cite{pabutools} to support project interaction, using it to evaluate methods on both pabulib instances and synthetic simulations. The experiment shows IES outperforms ES with substitute interactions. However, PIES, while ensuring proportionality for any interactions, results in higher welfare variance.

% We extended pabutools~\cite{pabutools} to support aggregation with the interaction between projects, which was used to evaluate the methods on different instances from pabulib in addition to synthetic simulations. The experiment shows that IES has a clear advantage over ES when having only substitute interactions. However, PIES is more complicated, guaranteeing proportionality for any interactions, but the cost of higher welfare variance.

% The experiment shows that IES and PIES achieve better welfare than ES (which does not take interactions into account) on average, in addition to having a higher number of instances where they achieve better welfare.

A natural follow-up is whether a polynomial-time aggregation rule can satisfy EJRI-1 for general interactions. Additionally, our work assumes a joint partition of projects for all voters, raising the question of whether a rule can ensure proportionality while allowing voters to create their own partitions for project interaction. Lastly, we used vote similarity to identify project interactions, which simplifies voting but may not be the best method. Future work could explore alternative ways to define interactions or test different elicitation methods that balance simplicity with accurate interaction representation.

% A natural follow-up question is whether there is a polynomial time aggregation rule that holds EJRI-1 for general interactions. Furthermore, our work assumes a joint partition of the projects used by all voters; thus, it raises the question of whether there exists a polynomial time aggregation rule that guarantees proportionality while allowing voters to create their own partition for the substitute projects. Finally, we used the similarity between votes to find project interactions. Its main advantage is that it keeps the votes simple (voters use approval voting); however, it is not necessarily the best way to define interactions. Therefore, one can either think about different methods to create interactions or test different elicitation methods with interactions to achieve a good representation of interactions while keeping it simple enough to use. 

%%%%%%%%%%%%%%%%%%%%%%%%%%%%%%%%%%%%%%%%%%%%%%%%%%%%%%%%%%%%%%%%%%%%%%%%

\begin{small}
\bibliographystyle{splncs04nat}

\bibliography{ref.bib}
\end{small}
%%%%%%%%%%%%%%%%%%%%%%%%%%%%%%%%%%%%%%%%%%%%%%%%%%%%%%%%%%%%%%%%%%%%%

\clearpage
\begin{subappendices}

\section{Omitted Proofs}\label{apx:proofs}

In section~\ref{sec:prop} we mention the notion of EJR-1 and EJRI-1, which are a relaxation of EJR and EJRI. The difference between EJR and EJR-1 or EJRI and EJRI-1 is whether or not we allow proportionality "up to 1 project" i.e. we might need to add 1 project to satisfy it. Next, we will show that the stronger notions hold that EJRI coincides with EJR under additive utilities and EJR implies EJRI under substitute interactions. % and FJR implies EJRI for any interactions.

\begin{proposition}\label{prop:ejr_to_ejri}
    EJR if exists $\rightarrow$ EJRI under substitute interactions.
\end{proposition}

To prove this and the following proposition, we will note by $\alpha_A$ and $\alpha_S$ as $\alpha$ functions defined by the additive and interaction definitions, respectively, i.e., whether it is defined per project or projects set.

\begin{proof}
    Suppose some aggregation method holds EJR (when such an outcome exists), and let's look at some $(\alpha_S,T)$-cohesive group $S$. For every $i\in S$, every project $p\in T$ and a subset of projects $B\subseteq P\setminus\{p\}$ it holds $MU_{\alpha_S}(p|B) = min_i u_i(p| B)$.

    We can define an additive $\alpha_A$ function such that $\alpha_A(p)=\alpha_S(p|\varnothing)=min_i u_i(p|\varnothing)$. Given $\alpha_A$ it holds that group $S$ is $(\alpha_A,T)$-cohesive (under additive definition) as $\forall p\in P; u_i(p|\varnothing)\geq\alpha_A(p)=min_i u_i(p|\varnothing)$. 

    Since we know that the method holds EJR, we have a voter $i\in S$ such that its outcome $B$ holds $u_i(B)\geq \alpha_A(T)\geq\alpha_S(T)$ as $\alpha_S$ can be only smaller when taking substitution into account. Therefore, it also holds EJRI.
\end{proof}

\begin{proposition}%\label{prop:ejr_to_ejri}
    EJRI $\leftrightarrow$ EJR under additive utilities.
\end{proposition}

\begin{proof}
    To prove this, we note that EJR $\rightarrow$ EJRI using the same proof as in Proposition~\ref{prop:ejr_to_ejri}.

    Next, we would like to show that EJRI $\rightarrow$ EJR under additive utilities.

    We start with the definition of cohesiveness, where the difference between EJR and EJRI is only the definition of $\alpha$.  
    Given a $(\alpha_A,T)$-cohesive group $S$, it holds that $u_i(p)\geq \alpha_A(p)$ for all $i\in S$ and $p \in T$. 
    Given a $(\alpha_S,T)$-cohesive group $S$, it holds that $MU_{\alpha_S}(p|B)= min_{i\in S}u_i(p|B)$ for all $p \in T$ and $B\subseteq P$.

    We first note that the requirement for $\alpha_A$ is equivalent to $u_i(p)\geq \alpha_A(p)=min_{i\in S}\alpha_A(p)$ since the requirement is for all voters.
    Next, as we are in the additive setting it holds $MU_{\alpha_S}(p|B) = MU_{\alpha_S}(p|\varnothing)=min_{i\in S}u_i(p|\varnothing)$. since $MU_{\alpha_S}(p|\varnothing)=\alpha_S(p|\varnothing)$ we get that that $\alpha_A$ and $\alpha_S$ coincides.

    Therefore, if any group of voters is $(\alpha_S,T)$-cohesive, then it is also $(\alpha_A,T)$-cohesive and the other way around.

    Since $\alpha_A = \alpha_S$ (due to additive utilities) it holds for any $(\alpha_A,T)$-cohesive group that $u_i(R(E))\geq \alpha_S(T)=\alpha_A(T)$
\end{proof}
The proof for the variations of up to one project is similar. 

In addition to the relation between EJR and EJRI, we would like to show the relation between the stronger notion FJR and EJRI.

\begin{proposition}\label{prop:FJR_to_EJR}
    FJR implies EJRI.
\end{proposition}

\begin{proof}
    Suppose that some aggregation method holds EJR and lets look at some $(\alpha,T)$-cohesive group $S$. For every $i\in S$, every project $p\in T=\{t_1,\ldots,t_{|T|}\}$ and a subset of projects $B\subseteq P\setminus\{p\}$ it holds $u_i(p| B)\geq MU_{\alpha}(p_ B)$. We set  $\beta=\alpha(T)$. Then the following holds: 
    $$u_i(T)=\sum_{j=1}^{|T|}u_i(t_j| \cup_{k=1}^j t_k)\geq\sum_{j=1}^{|T|}MU_{\alpha}(t_j, \cup_{k=1}^j t_k)=\alpha(T)$$
    Thus $S$ is weakly $(\beta,T)$-cohesive. As the method hold FJR we have a voter $i\in S$ such that its outcome $W$ holds $u_i(W)\geq\alpha(T)$, as required.
\end{proof}

\section{Omitted Examples}\label{apx:example}

In section~\ref{sec:prop} we saw that there isn't always an outcome that satisfies EJR-1 under the interaction settings. Example~\ref{exmp:ejr_failure} give a detailed example for such a scenario.

\begin{example}[EJR-1 in the interaction settings]\label{exmp:ejr_failure}
    Given PB instance with 8 projects split to partition: $z_1=\{p_1,\ldots,p_4\}, z_2=\{p_5,\ldots,p_8\}$. There is a single voter with a budget of 4.

    In addition, the projects in $z_1$ are total substitutes i.e. the first worth 1 and the rest 0, each cost 1. All projects in $z_2$ worth 1 and cost $1+\epsilon$. 

    It is easy to see that the single voter is $(\alpha,z_1)$-cohesive for $\alpha\equiv 1$, so $\alpha(z_1)=4$, therefore EJR-1 require that either $u_i(W) \geq \alpha(z_1)$ or  there is some project $p^*\in z_1$ such that $u_i(W\cup\{p^*\}) > \alpha(z_1)$.

    There are three cases for $W$:
    \begin{itemize}
        \item $W$  contains at no projects from $z_1$. Then  $W$ contains at most 3 projects so  even with an additional project $p^*$ we have $u_i(W)<4$ and $u_i(W\cup \{p^*\})\leq 4$.
        \item $W$ contains at least one project from $z_1$ but not all. Then $W$ is still blocked by $z_1$, as $u_i(W\cup \{p^*\})=u_i(W)\leq 3<\alpha(z_1)$.
        \item $W=z_1$. Then  $u_i(W)=1 < \alpha(z_1) = 4$.

    \end{itemize}

    The last case demonstrates a big issue where an outcome can block itself as $\alpha$ only consider singletons while the utility affected from projects interaction. Even if we disallow this, in the last case we have $T=\{p_5,p_6,p_7\}$. Our single voter is $(\alpha,T)$-cohesive for $\alpha\equiv 1$, and $u_i(W)=1 < \alpha(T) = 3$.
    
\end{example}

While this example show a case where EJR-1 outcome does not always exist, we can do a simple modification to show that EJR-1 and EJRI-1 are in-comparable. consider the same example, but the utility for all projects in $z_1$ are 1, and the marginal utilities for projects in $z_2$ grows exponentially (with base 10). This time, the only outcome to satisfy EJR-1 is $z_1$ ($z_1$ guarantee utility of 4 and $z_2$ guarantee utility of 3 according to EJR-1) while EJRI-1 is satisfied by choosing 3 projects from $z_2$, guaranteeing utility of 111. As can be seen different outcomes satisfy EJR-1 and EJRI-1 without any overlapping.
%%%%%%%%%%%%%%%%%%%%%%%%%%%%%%%%%%%%%%%%%%%%%%%%%%%%%%

\section{Omitted Definitions}

\begin{definition}[Fully Justified Representation (FJR)~\cite{peters2021proportional}] \label{def:fjr}
A group $S\subseteq V$ of voters is weakly $(\beta,T)$-cohesive for $\beta\in\mathbb{R}$ and a set of projects $T\subseteq P$, if $|S|/n \geq cost(T)/L$ and $u_i(T) \geq \beta$ for every voter $i \in S$.

An aggregation method $\mathcal{R}$ satisfies FJR if for each instance $E$ and each weakly $(\beta,T)$-cohesive groups of voters $S$ there exists a voter $i\in S$ such that $u_i(\mathcal{R}(E))\geq\beta$
\end{definition}

%%%%%%%%%%%%%%%%%%%%%%%%%%%%%%%%%%%%%%%%%%%%%%%%%%%%%%

\section{Equal Shares Aggregation}\label{apx:agg}

In Section~\ref{sec:agg}, we introduced ES and IES and explained how they work. In this section, we will present a pseudo-code for them.

Algorithm~\ref{algo:ES_agg} presents the pseudo-code for both ES and IES where $b_i(t)$ represents the funds left to voter $i$ at iteration $t$, and $B_t$ is the projects chosen up to iteration $t$. Both methods have the same high-level behavior, i.e., finding the projects with the smallest qValue and adding it to the chosen outcome.

The difference between the two methods appears in Algorithm~\ref{algo:qval}. In order to calculate the qValue of some projects, we need to consider the utility that each voter gets for this project. Therefore, we consider the projects chosen so far and look at the utility accordingly, i.e., $u_i(p|B_t)$. We note that if we are in the additive settings, the voter's utility is unaffected by $B_t$ and depends only on the utility for $p$.

\begin{algorithm}
\SetAlgoLined
\textbf{Input:}
PB instance $(P,cost,\mathcal{Z},L,V, F)$

\KwResult{Feasible bundle $B\subseteq A$}
 $B_0 \leftarrow \varnothing$

 $\forall i\in V: b_i(0) \leftarrow\frac{L}{|V|}$ 
  $t \leftarrow 1$

 \While{True}{

    % \HiLi$\forall i\in V, \forall p\in A: U_i(p) \leftarrow u_i(B_{t-1} \cup \{p\}) - u_i(B_{t-1})$ ~~~~// only in SRX
 
  $p^{(t)} \leftarrow argmin_{p\in P\setminus B_{t-1}}[qValue(p, u_{[|V|]}),B_{t-1}]$
  
  \If{$qValue(p^{(t)}, u_{[|V|]}),B_{t-1})=\infty$}{
        $return \  B_{t-1}$ 
   }
   $B_{t} \leftarrow B_{t-1} \cup \{p^{(t)}\}$
   
   $\forall i\in V: b_i(t) \leftarrow \max\{0,b_i(t-1) -U_i(p^{(t)})\cdot q\}$
   
   $t \leftarrow t + 1$
   }
 \caption{(Interaction) Equal Shares}\label{algo:ES_agg}
\end{algorithm}

\begin{algorithm}[t]
\SetAlgoLined
\textbf{Input:} 
\begin{enumerate}
    \item project $p\in P$ \\
    \item the set of chosen projects $B_t$
    \item $\forall i\in V,  u_i(p|B_t)$ \\
    
\end{enumerate}
\KwResult{q-value computation for project $p$}
\If {$\sum_{i\in v, u_i(p|B_t)>0}b_i(t)< cost(p)$}
 {$return \  \infty$}
 
 $current\_utility \leftarrow \sum_{i\in V}u_i(p|B_t)$
 
 $cost\_leftover \leftarrow cost(p)$
 
 $removed\_voters \leftarrow \varnothing$
 
 \While{True}{
    $current\_q \leftarrow cost\_leftover /    current\_utility$
    
    $voter\_removed \leftarrow False$
    
    \For{$i\in V\setminus removed\_voters$}{
      \If{$current\_q * u_i(p|B_t) > b_i(t)$}{
            $current\_utility \leftarrow current\_utility - u_i(p|B_t)$
            
            $cost\_leftover \leftarrow cost\_leftover - b_i(t)$
            
            $removed\_voters \leftarrow removed\_voters \cup \{i\}$
            
            $voter\_removed \leftarrow True$
       }
       
     }
     
     \If{$voter\_removed == False$}{
            $return \  (cost\_leftover / current\_utility)$
       }
 }

 \caption{qValue}\label{algo:qval}
\end{algorithm}

\section{Partition Interaction Equal Shares}\label{apx:pies}

In section~\ref{sec:prop} we saw that IES holds EJRI-1 for substitute projects, however this does not hold anymore for general interactions. For this reason, we suggest a variation of IES  called \emph{Partition Interaction Equal Shares} (PIES). This mechanism is same as IES, but with a preprocessing step:  for each part $z\in Z$ and every $T\subseteq Z$, add a new project $p_T$ and remove all original projects. We set $cost(p_T) := cost(T)$ and $u_i(p_T):= u_i(T)$ for all $i\in V$. %creating a new project for each subset $T\subseteq z$ of projects having the cost as the total cost of the projects that created it, and its marginal utility defined as the total marginal utility of the projects that created it. 
Once we have the new set of projects the aggregation will be performed similarly to IES, with the difference that at the end of each iteration where a project $p_T$ is chosen, all projects $p_{T'}$ with $T\cap T'\neq \emptyset$ are removed. 

When running PIES we have a larger amount of projects which is exponential in $|z^*|:=\max_{z\in Z}|z|$. This means that at each iteration we need to iterate over $O(|Z|2^{|z^*|})$ projects instead of only $O(M)$, making PIES less efficient compared to ES and IES. However, it is likely to assume that each interaction set size is bounded by some relatively small value, which in this case we get that the aggregation will still run in reasonable time.

\begin{proposition}
    PIES holds EJRI-z for any interaction function.
\end{proposition}

We give intuition for the proof. When using PIES, we look at a PB instance where every combination of projects in each interaction set is represented as a project. In this scenario, all interaction functions actually behave as functions for substitute projects, this due to the fact that choosing a project will remove all other overlapping projects and the utility for any non-overlapping project must be lower otherwise PIES would have chosen the project that represent both projects. PIES behave similarly to IES which we shown to hold EJRI-1 when all interaction functions are proportional, but since in PIES case each project can represent several projects from the same interaction set, the "up to 1 project" becomes "up to 1 interaction set".

\section{Proportionality with Multiple Partitions}\label{apx:multi_part}

In Section~\ref{sec:prop} we saw that IES and PIES can achieve proportionality in our settings. In this section, we consider the scenario where each voter can submit a different partition, demonstrating that those methods fail to satisfy proportionality.

\begin{example}
    Given a participatory budgeting scenario with 3 voters $\{v_1,v_2,v_3\}$ and 16 projects $\{a_{1-3},b_{1-3},c_{1-3},d_{1-3},e_{1-4}\}$, where $cost(a_{1-3})=cost(b_{1-3})=cost(c_{1-3})=cost(d_{1-3})=1, cost(e_{1-4})=\frac{3}{2}$ and for each $i\in[1,3]$ the voters approve the following:
\begin{itemize}
    \item $v_1: \{(b_i),(c_i),(a_i,d_i)\}$
    \item $v_2: \{(b_i),(a_i,c_i),(d_i)\}$
    \item $v_3: \{(a_i,b_i),(c_i),(d_i)\}$
\end{itemize}
Where projects at the same parenthesis the voters want exactly one of them (the second project will have utility of zero). In addition, all voters approve $e_{1-4}$ (without interaction). 

The utility for all projects (expect for the interactions) is one and the total budget is $L=9$. We will use IES (PIES works similarly)  for aggregation with tie-breaking for the worst case (tie-breaking done for easier readability, it is possible to change the utility or cost by a small $\epsilon$ value and the outcome will not change without the need for tie-breaking).

At the first step projects $a_{1-3},b_{1-3},c_{1-3},d_{1-3}$ are $\frac{1}{3}$-affordable, while projects $e_{1-4}$ are $\frac{1}{2}$-affordable. Using tie breaking, IES will choose to fund project $a_1$, leaving each voter with budget of $\frac{8}{3}$.

After choosing project $a_1$, the utility of the voters update accordingly $u_1(d_1)=u_2(c_1)=u_3(b_1)=0$, resulting that projects a,b,c becoming $\frac{1}{2}$-affordable, while the other projects remain the same. In addition, each voter have a budget of two remaining.

In similar manner, projects $a_2$ and $a_3$ will be chosen. This results with all projects being $\frac{1}{2}$-affordable as only two voters give utility > 0 for projects $b_{1-3},c_{1-3},d_{1-3}$, while the utility of $e_{1-4}$ projects is still unchanged.

As there are not any interactions left, the utility for all projects will stay the same until the aggregation stops and projects will be chosen by tie-breaking. For this reason, we tie-break in favor of $e_{1-4}$ and choosing to fund all of them, which result with exhausting the budgeting and getting an outcome of $W = \{a_{1-3}, e_{1-4}\}$ which give utility of 7.

Lets note the projects set $T=\{a_{1-3},b_{1-3},c_{1-3}\}$ and set of voters $S=\{v_1,v_2,v_3\}$ which are T-cohesive with $\alpha(c,B_t)=1$, therefore there is at least one voter which should get utility of 9. However, the outcome utility is 7 and even when adding one more project will be 8, still lower than required. This is a violation of EJR-1.
\end{example}

An intuition to why IES does not hold EJR-1 in the example, is looking at the projects $a_i - d_1$ as a single interaction set where the interaction function can give a different value for different sets of projects and does not look only at amount i.e. it isn't indifferent to the which project is funded anymore. Next, note that the guarantee of EJR-1 for $(\alpha,T)-cohesive$ groups depends only by the utility of projects in $T$, however if the interaction functions is not indifferent to which project is chosen, there might be some project outside of $T$ which hurts the utility of $T$.

\section{Welfare Worse-Case Analysis}\label{apx:welfare}

In section~\ref{sec:worst} we saw the welfare ratio of ES in the additive settings, here we will dive into the reason we get such ratio and demonstrate it with an example.

When running ES (or one of its variations), in each iteration we search for the project that currently has the lowest qValue, however while it is practically calculated as the ratio between cost and utility, it does not necessarily mean that projects with high utility will be chosen. The reason for this is that the qValue should also be ``fair", which is reflected by higher qValue when supporters' budget is constrained. % When the a project is selected, the supporters should pay proportionally to how much utility they ``donated" for the project, however if some of them do not have enough funds to pay proportionally they need to use their entire funds and the rest of the voters pay more than their proportional utility. To handle this issue, the qValue takes this into account and increase according to those voters who need to pay more (This formally happen in Equation~\ref{eq:afford} by taking $b_i$ for the voters that does not have enough funds when calculating the qValue).

This behavior mean that ES variations will prioritize  projects where there is larger agreement about how much utility this project is worth i.e. projects with lower variance over their utility. Therefore, ES can result with outcomes with worse social welfare to maintain the fairness. This downside can become more noticeable when using PIES. The reason for this that when considering a set of projects it might have a high utility for many voters (including complementary projects), but even one voter which does not agree with it and give it a low utility it will have a very high qValue. Therefore, will miss the option to choose projects with very high utility.

To demonstrate this behavior, lets look at a scenario with 5 voters, 2 projects that cost 10 and budget of 10. All of the voters approve the first project with utility of 2 and the second project with utility 10 except for one voter which give it utility of 2 (note we use additive utilities). The first project will be 1-affordable, while the second project with much higher welfare will be 2-affordable. In both cases all 5 voters will use their entire funds, however the project with higher total utility will have higher qValue, thus the other project will be funded. This behavior will be further demonstrated in experiment at Section~\ref{sec:exp}.

This scenario can be further extended such that the utility of the second project is increased from 10. No matter how much we increase this value, as long there is one voter with low utility, the first project will still be chosen resulting with the same welfare while the optimal social welfare keep increasing.

\section{Simulations Full Results}\label{apx:sim}

\begin{figure}[h!]
\begin{center}
\includegraphics[width=8cm]{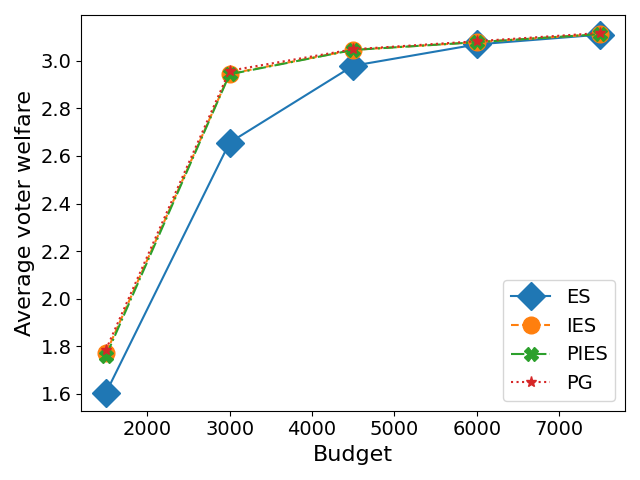}
\caption{Average voter welfare over 1000 simulations, given only minimal substitutes voters.
}\label{fig:type1_app}
\end{center}
\end{figure}

\begin{figure}[bh!]
\begin{center}
\includegraphics[width=8cm]{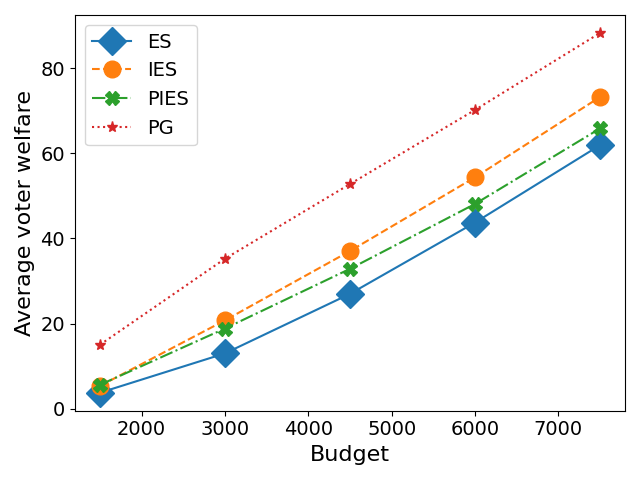}
\caption{Average voter welfare over 1000 simulations, voters have mix of complementary interactions i.e. extreme complementary, linear and exponential comp .
}\label{fig:mix_comp}
\end{center}
\end{figure}

\begin{figure}[bh!]
\begin{center}
\includegraphics[width=8cm]{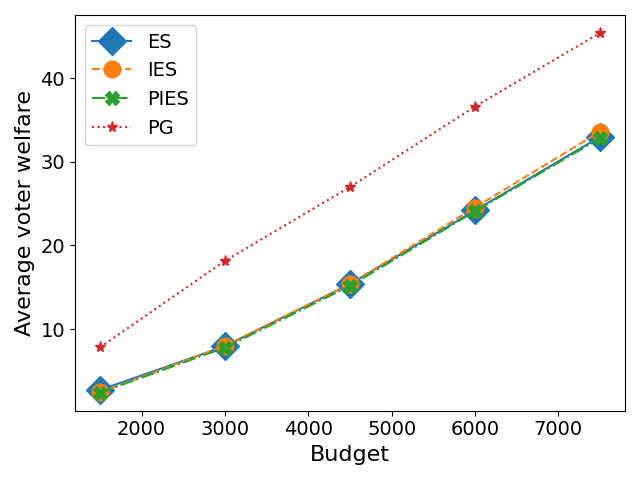}
\caption{Average voter welfare over 1000 simulations, voters have mix of all possible interactions.
}\label{fig:mix_all_app}
\end{center}
\end{figure}

This section presents the results from the experiments described in Section~\ref{sec:exp} for all types of interaction functions in Figures~\ref{fig:type1_app}-\ref{fig:mix_all_app}. We remind the reader of the six types of interaction functions:

\begin{enumerate}
    \item Type 1 (minimal substitutes) - the voter prefers one non-substitute project over the entire set of substitute projects i.e. the utility for the first project is 1, while any additional project gives the utility of $1/m$

    \item Type 2 (harmonic) - the marginal utility decrease by harmonic series, formally: $f(i)=\sum_{j=1}^i\frac{1}{j}$

    \item Type 3 (exponential sub) - the marginal utility decreases exponentially, formally: $f(i)=\sum_{j=1}^i \frac{1}{1.5^{j-1}}$

    \item Type 4 (extreme complementary) - the voter significantly prefer at least two projects and want as many as possible, formally the marginal utility for the first project is 1 and 50 for the rest.

    \item Type 5 (linear) - the marginal utility increase linearly, formally: $f(i) = \sum_{j=1}^i j$

    \item Type 6 (exponential comp) - the marginal utility increases exponentially, formally: $f(i)=\sum_{j=1}^i 1.5^{j-1}$
\end{enumerate}

Similarly to Section~\ref{sec:exp}, we see in Figures~\ref{fig:type1_app}-\ref{fig:mix_comp} that IES and PIES succeed in achieving better welfare compared to ES. However, in some scenarios such as Figure~\ref{fig:mix_comp}, due to the high variance in voter utilities for projects, PIES may prioritize interaction sets with lower welfare, resulting in lower welfare than IES.

Finally, in Figure~\ref{fig:mix_all_app} where voters can utilize all interaction types, we observe the behavior outlined earlier. With a limited budget, IES/PIES lacks flexibility for exploration, potentially leading to suboptimal choices and slightly lower welfare compared to ES. However, as the budget expands, the gap diminishes until IES begins to surpass ES in welfare again (with PIES maintaining welfare close to ES).

\begin{figure}[h!]
\begin{center}
\includegraphics[width=8cm]{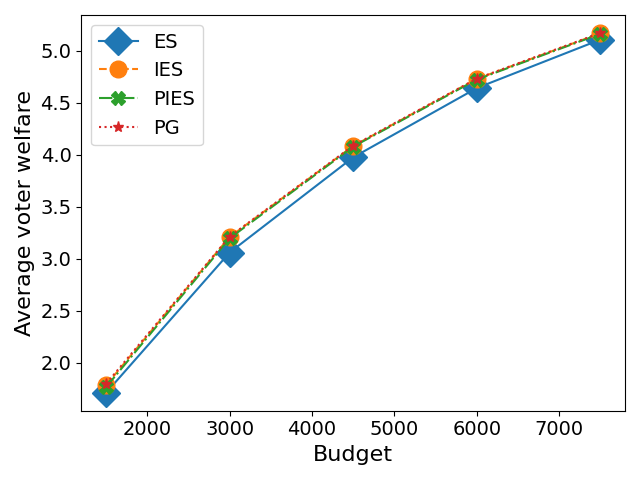}
\caption{Average voter welfare over 1000 simulations, given only harmonic voters.
}
\end{center}
\end{figure}

\begin{figure}[h!]
\begin{center}
\includegraphics[width=8cm]{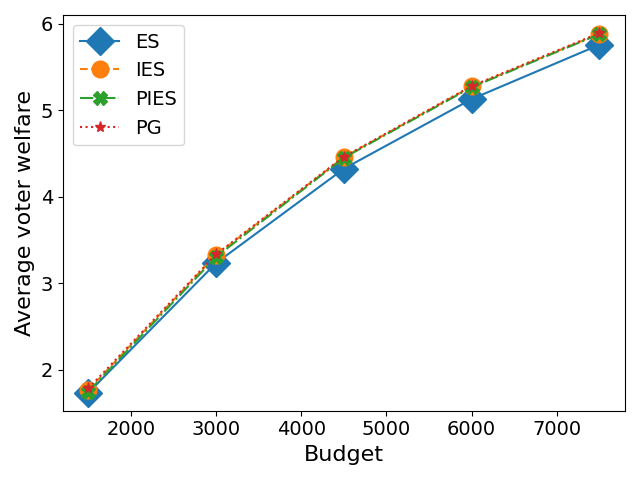}
\caption{Average voter welfare over 1000 simulations, given only exponential sub voters.
}
\end{center}
\end{figure}

\begin{figure}[h!]
\begin{center}
\includegraphics[width=8cm]{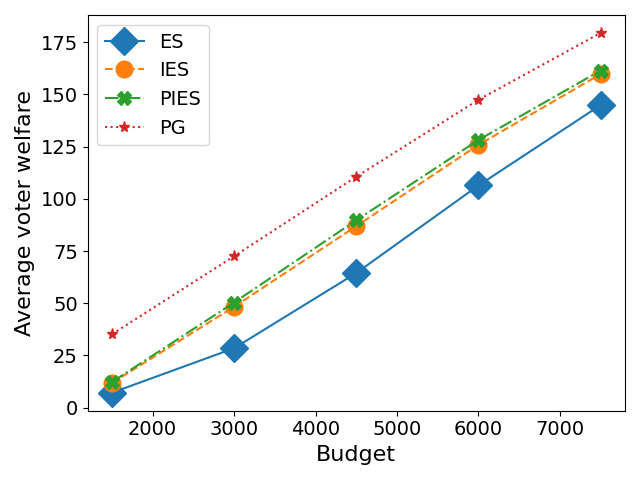}
\caption{Average voter welfare over 1000 simulations, given only extreme complementary voters.
}
\end{center}
\end{figure}

\begin{figure}[h!]
\begin{center}
\includegraphics[width=8cm]{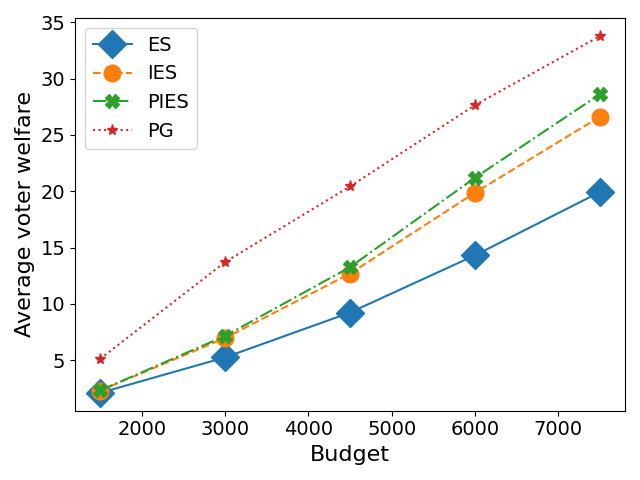}
\caption{Average voter welfare over 1000 simulations, given only linear voters.
}
\end{center}
\end{figure}

\begin{figure}[th!]
\begin{center}
\includegraphics[width=8cm]{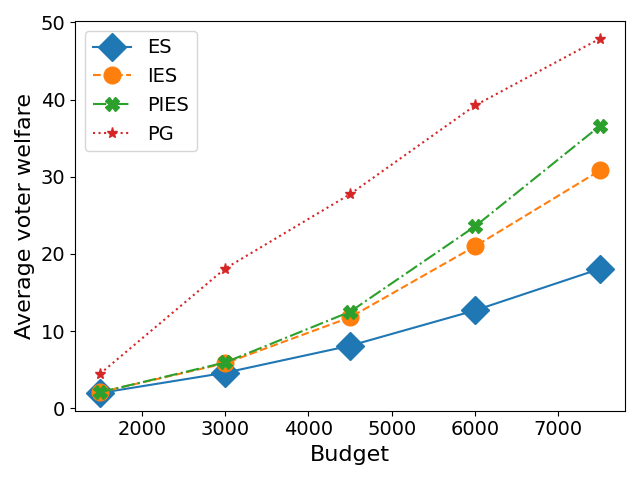}
\caption{Average voter welfare over 1000 simulations, given only exponential comp voters.
}
\end{center}
\end{figure}

\begin{figure}[bh!]
\begin{center}
\includegraphics[width=8cm]{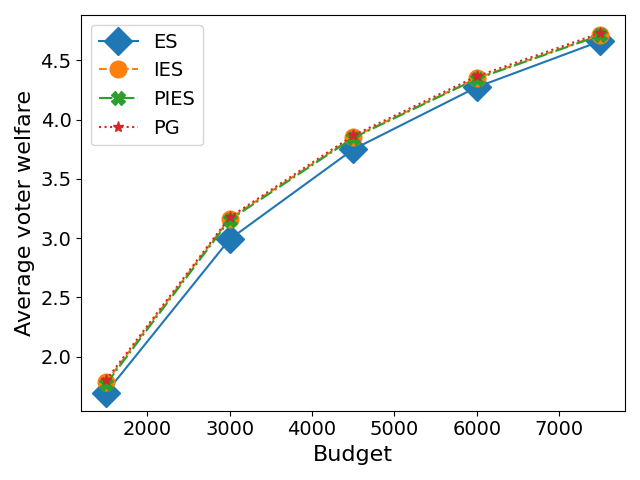}
\caption{Average voter welfare over 1000 simulations, voters have mix of substitution interactions i.e. 
minimal substitutes, harmonic and exponential sub. }
\end{center}
\end{figure}

\end{subappendices}

\end{document}